\documentclass[a4paper,11pt]{article}
\pdfoutput=1
\usepackage{jheppub}
\usepackage[T1]{fontenc}
\usepackage{amsmath}
\usepackage{amssymb}
\usepackage{mathrsfs}
\usepackage{graphicx}
\usepackage{mathtools}
\usepackage{subcaption}
\usepackage{booktabs}
\usepackage{makecell}

\def\be{\begin{equation}}
\def\ee{\end{equation}}
\def\e{\text{e}}

\def\dd{\text{d}}

\def\max{\text{max}}

\def\V{\mathcal{V}}

\usepackage{changes}
\definechangesauthor[name={Per cusse}, color=orange]{per}

\begin{document}
\title{Constraints on Kasner Exponents from Holography and Energy Conditions}

\author{Zi-Hao Li, Run-Qiu Yang}
\affiliation{Center for Joint Quantum Studies and Department of Physics, School of Science, Tianjin University, Yaguan Road 135, Jinnan District, 300350 Tianjin, P.~R.~China}
\emailAdd{lieaction.lzh@gmail.com}
\emailAdd{aqiu@tju.edu.cn}
\abstract{A central question in bottom-up holography is whether a given bulk effective theory admits a consistent holographic dual. In this work, we explore whether the near-singularity Kasner scaling of planar AdS black hole interiors can serve as a useful diagnostic based on interior-sensitive holographic probes. By examining the metric combinations that control interior-sensitive observables, including Complexity=Volume (CV), the bulk contribution to Complexity=Action (CA), Hartman-Maldacena (HM) entropy, and the thermal $a$-function, we derive algebraic bounds on the Kasner exponents $(p_t,p_s)$ for the relevant semi-classical regimes: finite-radius late-time branches of CV and HM, finite CA complexity, and a finite near-singularity thermal $a$-function. The resulting inequalities delineate the corresponding regions of Kasner space in which the terminal scaling alone is sufficient to realize these behaviors. Furthermore, we demonstrate that classical energy conditions, specifically the null and dominant energy conditions, provide simple sufficient criteria ensuring that the Kasner exponents fall within these holographically allowed regions. These results establish a direct connection between classical bulk energy conditions and the interior geometry selected by holographic probes, and suggest that Kasner scaling can provide a complementary diagnostic in bottom-up holography.}

\maketitle
\flushbottom

\section{Introduction}\label{sec:intro}

Holographic duality establishes an equivalence between a bulk gravitational theory and a lower-dimensional quantum field theory living on its boundary~\cite{Maldacena:1997re, Gubser:1998bc, Witten:1998qj, Maldacena:2001kr, Maldacena:2013xja}. Given this framework, a fundamental question arises: which bulk gravitational theories actually admit a consistent boundary dual? In top-down constructions, this question is answered by embedding the bulk theory into a UV-complete theory of quantum gravity, such as string theory. For most effective bulk theories, however, no such embedding is known.
In practice, one therefore adopts a bottom-up approach, starting from an effective bulk theory to model strongly coupled boundary dynamics. This pragmatic approach has been extremely fruitful, but it also raises a consistency question: not every reasonable bulk effective theory will define a consistent holographic CFT.

Existing consistency tests for candidate holographic theories have often been formulated in terms of boundary causality and microcausality constraints~\cite{Brigante:2008gz, Camanho:2009vw, Camanho:2014apa}, conformal-collider or energy-flux positivity bounds~\cite{Hofman:2008ar, Camanho:2009vw}, restrictions on CFT anomaly coefficients and holographic RG flows~\cite{Henningson:1998gx, Freedman:1999gp, Myers:2010tj}, as well as consistency conditions for holographic entanglement entropy~\cite{Headrick:2007km, Wall:2012uf, Headrick:2014cta}. These examples show that holography itself can act not only as a dictionary for computing boundary observables, but also as a consistency principle constraining the space of admissible bulk effective theories.

Holographic observables can also probe regions deep into the interior of black holes. Several geometric prescriptions are sensitive to regions behind the event horizon. Examples include Hartman-Maldacena (HM) extremal surfaces in two-sided black holes~\cite{Ryu:2006bv, Hubeny:2007xt, Hartman:2013qma},  holographic complexity (the Complexity=Volume (CV)~\cite{Susskind:2014rva, Stanford:2014jda} and Complexity=Action (CA) conjectures~\cite{Brown:2015bva, Brown:2015lvg, Lehner:2016vdi}), and the thermal $a$-function~\cite{Caceres:2022smh, Caceres:2022hei, Caceres:2023zft}. Although these prescriptions probe the interior in different ways, each depends on specific combinations of the bulk metric, which naturally leads to the following question: can the interior requirements associated with these holographic constructions provide a useful diagnostic of holographic consistency for candidate bulk theories?

In this work, we answer this question by exploring a complementary diagnostic based on black hole interiors. We ask how the terminal Kasner scaling is constrained by the interior requirements of several holographic probes. For each construction, we first identify the geometric or endpoint property required by its standard semi-classical regime, and then ask when the terminal Kasner scaling is sufficient to guarantee that property.
A wide range of holographic black holes with matter hair develop a space-like singularity governed by a terminal Kasner regime. This universal behavior typically emerges when scalar kinetic terms dominate the asymptotic interior dynamics~\cite{Cai:2020wrp} or through cosmological-billiard mechanics for massive fields~\cite{Henneaux:2022ijt}. Prominent examples include homogeneous relevant deformations of thermal CFTs~\cite{Frenkel:2020ysx,Wang:2020nkd}, holographic superconductors~\cite{Hartnoll:2020fhc,Sword:2021pfm}, and explicit Einstein-Maxwell-dilaton models~\cite{Arean:2024pzo}. Kasner singularities have also been studied in asymptotically flat Einstein-scalar black holes, including their relation to asymptotic mass and scalar-charge data~\cite{Xiong:2026npi}. Within these geometries, the associated Kasner exponents capture rich interior dynamics, exhibiting characteristic jumps across holographic phase transitions~\cite{Liu:2021hap,Gao:2026tck} or supporting transitions between distinct Kasner epochs~\cite{Zhang:2025hkb,Zhang:2025tsa}. 

While these interior geometries can exhibit complex multi-epoch dynamics, our analysis focuses on static and transversely isotropic backgrounds that ultimately settle into a single, constant-exponent Kasner regime. In this symmetric setup, only two independent exponents, $(p_t, p_s)$, are required to characterize the Kasner singularity.
Consequently, we exclude interiors that never reach a final Kasner regime, such as those with persistent BKL evolution~\cite{Belinsky:1970ew,Belinsky:1982pk}, never-ending Kasner alternations~\cite{An:2022lvo}, non-Kasner asymptotic epochs~\cite{Cai:2023igv}, and singularities dominated by higher-derivative dynamics~\cite{Caceres:2024edr}. We also exclude geometries that terminate at an inner Cauchy horizon rather than a space-like Kasner singularity~\cite{Poisson:1990eh,Cai:2020wrp}. Within this scope, the Kasner exponents provide a compact description of the leading metric scaling relevant to the holographic quantities studied below.

Our proposal is to use the geometric and endpoint requirements associated with these interior-sensitive holographic constructions to constrain the terminal Kasner data\footnote{An earlier relation between Kasner data, energy conditions, and holographic bounds was noted in a cosmological setting in Ref.~\cite{Cataldo:2001bn}, where a holographic entropy bound for Bianchi-I cosmologies was found to impose the same constraint on the scale-factor parameters as the dominant energy condition. The holographic quantities considered here are instead AdS/CFT probes of black-hole interiors.}. In a holographic dictionary, consistency with the boundary field theory suggests certain requirements on the black hole interior. For the geometric probes (CV, HM, and the thermal $a$-function), these requirements take the form of kinematic endpoint criteria; for the CA bulk action, they impose a non-divergence condition tied to the on-shell bulk action near the singularity. For each observable, we translate these requirements into a simple algebraic inequality on $(p_t, p_s)$ by demanding that the terminal geometry alone can guarantee the corresponding well-defined phase.
The resulting inequalities therefore provide sufficient Kasner criteria for the corresponding probe properties. The CV, HM, and thermal $a$-function bounds are purely kinematic, following directly from the asymptotic metric scaling, whereas the CA bound is dynamical and additionally relies on the Einstein equations and a restriction on the matter sector. For CV and HM, violation of the bound means that the endpoint argument no longer guarantees the existence of the desired finite-radius interior extremum. In such cases, the actual late-time behavior depends on the full pre-asymptotic radial profile rather than on the terminal Kasner scaling alone. The thermal $a$-function behaves differently because it probes the terminal region directly: violation of its Kasner bound leads to a divergent $a_T$ at the singularity. Likewise, if the CA inequality is violated, the singularity-side bulk contribution diverges, rendering the CA complexity ill-defined within the class of geometries considered here.

Since an arbitrary effective theory is not guaranteed to satisfy these constraints, we seek a more direct diagnostic: what fundamental properties of a bulk theory can ensure that these Kasner bounds are satisfied? After deriving these holographic Kasner bounds, we demonstrate that classical energy conditions provide a sufficient criterion for this purpose.
Specifically, the null energy condition (NEC) protects the bounds associated with the CV, HM, and thermal $a$-function observables, while the dominant energy condition (DEC) ensures the regularity of the CA bulk action. Thus, if a bulk theory satisfies the relevant energy conditions, its Kasner exponents automatically lie within the holographically allowed region.  Thus the energy conditions provide an independent dynamical route to the Kasner regions selected by the holographic probes.

Strictly speaking, resolving a space-like singularity requires a full theory of quantum gravity, which raises a natural concern about the validity of our bounds derived from classical gravity. Nonetheless, our semi-classical framework yields physically robust diagnostics by cleanly distinguishing probe-dependent sensitivities.
First, the geometric requirements relevant to the standard late-time behavior of CV and HM depend on an intermediate region rather than on the exact singularity. Although we assume the energy conditions hold throughout the entire spacetime for simplicity, our method can be generalized to cases where they are satisfied only in an intermediate region that does not reach the singularity. The relevant late-time extremal surfaces do not extend to the singular endpoint. At large boundary time, their deepest radial points approach fixed positions inside the horizon at a finite distance from the singularity. Hence our conclusions hold provided that semi-classical physics (Einstein equations and classical energy conditions) remains valid there. Second, classical energy conditions establish a finite ``buffer zone'' for these geometric probes: in the dimensions considered here, the energy-condition region lies a finite distance inside the CV- and HM-allowed regions. Even if subleading quantum effects mildly violate energy conditions~\cite{Martin-Moruno:2017exc, Kontou:2020bta} deeper in the interior, this buffer protects the critical surfaces from destruction.
By contrast, the bounds for the CA bulk action and the thermal $a$-function are marginal; they coincide exactly with the energy-condition boundaries without a buffer. Consequently, while highly sensitive to deep-interior quantum corrections, these observables serve as sharp boundary indicators for the breakdown of classical description. Together, these features ensure that the resulting Kasner bounds offer concrete, prescription-dependent semi-classical diagnostics for evaluating holographic consistency.

The remainder of this paper is organized as follows. In Sec.~\ref{sec:Kasner}, we introduce the asymptotically AdS black hole geometry and relate its near-singularity scaling to the Kasner exponents. In Sec.~\ref{sec:HolographicBound}, we derive Kasner bounds by identifying the geometric or endpoint conditions associated with several interior-sensitive holographic probes. In Sec.~\ref{sec:EnergyCondition}, we evaluate the classical null and dominant energy conditions and demonstrate how they serve as sufficient criteria for the holographic bounds. In Sec.~\ref{sec:discussion}, we summarize our results and discuss their future implications. Finally, Appendix~\ref{app:CABulk} collects the causal-structure, matter-sector, and on-shell reduction details used in deriving the CA bulk criterion. Appendix~\ref{app:KasnerEnergy} provides an independent geometric derivation of the energy-condition bounds using the orthonormal frame of a general Kasner cosmology.  

\section{Kasner Interiors of Asymptotically AdS Black Holes}\label{sec:Kasner}

\subsection{Geometric setup}

In the classical regime of holography, a homogeneous thermal state of a $d$-dimensional boundary theory is dual to a $(d+1)$-dimensional asymptotically anti-de Sitter (AdS) planar black brane. Working in $d\geqslant 3$ dimensions and setting the AdS radius to unity, we consider the following static planar metric:
\begin{equation}\label{eq:AdSmetric}
    \dd s^2  = \frac{1}{z^2} \left( -f(z)\e^{-\chi(z)}\dd t^2 + f^{-1}(z)\dd z^2 + \sum_{i=1}^{d-1} \dd x_i^2 \right)\ .
\end{equation}
The asymptotic boundary lies at $z=0$, where $f(z)\to 1$ and $\chi(z)\to 0$. The event horizon is located at $z=z_h$, where $f(z_h)=0$. In the region $0<z<z_h$, $f(z)>0$. Inside the horizon $z>z_h$, the blackening factor $f(z)$ becomes negative, causing the radial coordinate $z$ to become time-like. We restrict attention to interiors that terminate at a space-like singularity at $z\to\infty$ and asymptotically settle into a single constant-exponent Kasner regime. Such final regimes occur in many holographic models with matter hair, including relevantly deformed black holes~\cite{Frenkel:2020ysx,Henneaux:2022ijt}, holographic superconductors~\cite{Hartnoll:2020fhc,Sword:2021pfm}, multi-scalar models~\cite{Zhang:2025hkb}, and analytically controlled hairy solutions~\cite{Arean:2024pzo}. We exclude instead geometries that never settle into a final Kasner regime, as well as interiors that terminate at an inner Cauchy horizon or develop a different terminal scaling.

In the radial gauge of Eq.~\eqref{eq:AdSmetric}, a constant-exponent Kasner regime is equivalently represented by a power-law behavior of $f(z)$ and a logarithmic behavior of $\chi(z)$. We therefore write, as $z\to\infty$,
\begin{equation}\label{eq:assumofchi-f}
    f(z) \sim -f_\infty z^{2(a+1)}\ , \qquad \chi(z) \sim -2b\ln z+\chi_\infty\ ,
\end{equation}
with $f_\infty >0$ and all subleading terms suppressed. The constants $f_\infty$ and $\chi_\infty$ merely rescale the local coordinates and do not affect the Kasner exponents, whereas the parameters $a$ and $b$ control the critical power-law scaling near the singularity and therefore determine the Kasner data.

To ensure this $z\to\infty$ boundary represents a genuine physical singularity defined by geodesic incompleteness, the proper time $\tau = \int^\infty_z \dd z'\,\sqrt{-g_{z'z'}}$ along a zero-momentum radial time-like geodesic must be finite.  A direct calculation gives $\tau  \sim f^{-1/2}_\infty\int^\infty \dd z'\,z'^{-(a+2)}$, so for this proper time to be finite, we require $a > -1$. Furthermore, evaluating the Kretschmann scalar $K = R_{\mu\nu\rho\sigma}R^{\mu\nu\rho\sigma}$ near $z \to \infty$ yields $K \sim \mathcal{K}(a,b;f_\infty)z^{4(a+1)}$. For $d\geqslant 3$ and $a > -1$, the prefactor is strictly positive\footnote{For the spacetime dimensions $d \geqslant 3$ considered in this work, explicit calculation shows that this prefactor can be: $\mathcal{K}(a,b;f_\infty) = 4f_\infty^2 \left[(a+b)^2(2a+b+1)^2 +(d-1)a^2 +(d-1)(a+b)^2 +\frac{(d-1)(d-2)}{2} \right]$. For $d\geqslant 3$, $f_\infty>0$, the term in brackets is strictly positive, thus the curvature singularity is genuinely divergent for any real Kasner parameters.} $\mathcal{K}(a,b;f_\infty)>0$, confirming a genuine curvature divergence rather than a coordinate artifact.

\subsection{Kasner Metric and Exponents}

By mapping the radial coordinate $z$ in Eq.~\eqref{eq:AdSmetric} to the proper time $\tau \sim f^{-1/2}_\infty z^{-(a+1)}$, the leading metric coefficients scale as
\begin{equation}
    g_{tt} \sim f_\infty\e^{-\chi_\infty}z^{2(a+b)} \propto \tau^{-\frac{2(a+b)}{a+1}}\ , \qquad g_{x_i x_i}  = z^{-2} \propto \tau^{\frac{2}{a+1}}\ .
\end{equation}
After constant re-scalings of $t$ and $x_i$, the leading-order metric near a space-like singularity therefore adopts the Kasner form~\cite{Belinski:1973zz, Kasner:1921zz}:
\begin{equation}\label{eq:Kasner}
    \dd s^{2} = -\dd \tau^{2} + \tau^{2p_{t}} \dd t^{2} + \tau^{2p_{s}} \left(\sum_{i=1}^{d-1}\dd x_i^2\right)\ .
\end{equation}
Here the singularity is located at proper time $\tau \to 0$, and the coordinate $t$, inherited from the boundary time, is a spatial Kasner direction inside the horizon, while $\tau$ is the physical time approaching the singularity. Exponents $p_t$, $p_s$ are the associated Kasner exponents\footnote{We use ``Kasner exponents'' in the generalized sense of power-law exponents characterizing the leading anisotropic geometry. In the presence of matter, they need not satisfy the vacuum Kasner relations $p_t+(d-1)p_s=1$, $p_t^2+(d-1)p_s^2=1$.}, which could be expressed directly in terms of the metric scaling parameters $a$ and $b$:
 \begin{equation}\label{eq:definitionp_sp_t}
    \begin{aligned}
        p_s &= \frac{1}{a+1}\ ,\qquad p_t = -\frac{a+b}{a+1}\ ;\\
        \implies a &=\frac{1}{p_s}-1\ ,\qquad  b=\frac{p_s-p_t-1}{p_s}\ .
    \end{aligned}
\end{equation}
As established in the previous subsection, a physical space-like singularity requires $a>-1$, which directly translates to the positivity condition
\begin{equation}
    p_s > 0\ .
\end{equation}

As a reference point, for the vacuum AdS-Schwarzschild planar black brane $f(z)=1-\left(\frac{z}{z_h}\right)^d$ and $\chi=0$, we have $(p_t,p_s)=\left(\frac{2-d}{d},\frac{2}{d}\right)$, which satisfies both vacuum Kasner relations. In the following sections, we treat $(p_t,p_s)$ as the fundamental interior data. We first determine how the terminal Kasner scaling constrains the geometric or endpoint behavior relevant to several interior-sensitive holographic constructions. We then compare the resulting Kasner regions with the constraints implied by bulk energy conditions.

\section{Kasner Bounds from Interior-Sensitive Holographic Observables}\label{sec:HolographicBound}

We now use the Kasner data introduced in Sec.~\ref{sec:Kasner}. Our goal is to study the interior conditions associated with the consistency of several holographic prescriptions that probe the black hole interior. Our strategy is to separate two questions. We first identify the necessary geometric or endpoint property required by the standard regime of each holographic construction. We then ask when the terminal Kasner scaling is sufficient to guarantee that property. In each holographic probe, a simple combination of the metric functions $f(z)$ and $\chi(z)$ captures the relevant interior behavior, but the role of the terminal Kasner scaling is different. 
For CV and HM, the standard semi-classical late-time branch is associated with an extremal surface whose turning point approaches a finite radius behind the horizon, while the corresponding growth rate approaches a constant. We therefore ask whether the endpoint behavior of the relevant turning-point function guarantees the existence of a finite-radius interior extremum. For the CA bulk contribution and the thermal $a$-function, the relevant quantities extend directly to the terminal interior, and their Kasner criteria instead follow from requiring the corresponding near-singularity quantities to remain finite.

The resulting conditions are algebraic inequalities on $(p_t,p_s)$. For CV and HM, they characterize endpoint behavior that ensures the existence of a finite-radius critical point. For the CA bulk term and the thermal $a$-function they control terminal quantities directly. These bounds should be understood as holographic consistency conditions in some sense. They do not assume any classical energy condition. Instead, they follow from the assumption that the corresponding holographic proposal can be applied to the bulk geometry. Energy conditions will enter only in Sec.~\ref{sec:EnergyCondition}, where we show that they provide sufficient criteria for these bounds.

\subsection{Complexity=Volume and Interior Critical Surfaces}\label{subsec:CV}

\begin{figure}[htbp]
 \begin{center}
   \includegraphics[width=1\textwidth]{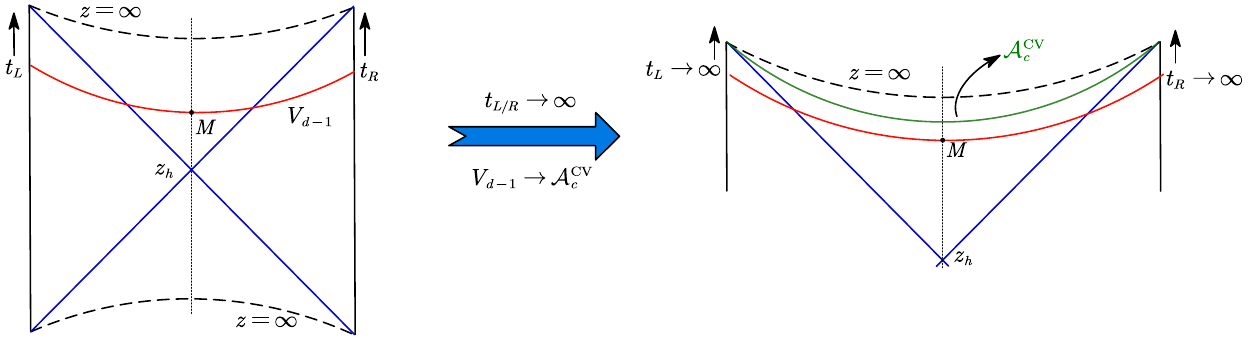}
 \end{center}
\caption{A maximal codimension-one slice connecting the two asymptotic boundaries at symmetric boundary times $t_L$ and $t_R$. The slice reaches its maximum radial depth at the turning point $M$, where $z=z_M$. Along the standard late-time branch, $z_M$ approaches a finite critical radius $z_{M,c}$. The associated limiting interior surface $\mathcal{A}_c^\text{CV}$ controls the asymptotic CV growth rate.
} \label{fig:CV-latetime}
\end{figure}

According to the Complexity=Volume (CV) conjecture~\cite{Susskind:2014rva, Stanford:2014jda}, the quantum complexity of a holographic boundary state at time $t$ is geometrically dual to the volume of the maximal codimension-one spatial slice $\Sigma$ in the bulk that is anchored at the boundary time slice, as shown in Fig.~\ref{fig:CV-latetime}. The complexity $C_V$ is defined as:
\begin{equation}
    C_V= \frac{\max(V_\text{CV}(\Sigma))}{G_\text{N} \ell_{\text{AdS}}}\ ,
\end{equation}
where $G_\text{N}$ is Newton's constant and $\ell_{\text{AdS}}$ is the AdS radius (which we set to unity, $G_\text{N}=\ell_{\text{AdS}}=1$, in our setup). Although the maximal volume itself contains the usual UV divergence near the asymptotic boundary, its time derivative is UV finite and provides a convenient quantity for studying the time evolution of CV complexity~\cite{Carmi:2017jqz}. In static eternal black holes, this growth rate approaches a constant along the familiar late-time branch. Geometrically, this behavior is associated with the turning point of the maximal slice approaching a finite radius behind the horizon. We now make this relation explicit and then ask what the terminal Kasner geometry can tell us about the existence of such a finite-radius extremum.

To evaluate the volume $V_\text{CV}(\Sigma)$ of the codimension-one surface that penetrates the black hole horizon smoothly, we introduce the Eddington-Finkelstein coordinate $v = t + z^*(z)$ and define $\dd z^*(z) \equiv f^{-1}(z)\e^{\chi(z)/2}\, \dd z$. Then in this Eddington coordinate the bulk metric~\eqref{eq:AdSmetric} is written as:
\begin{equation}
     \dd s^2 = \frac{1}{z^2} \left( -f(z)\e^{-\chi(z)}\dd v^2 + 2\e^{-\chi(z)/2}\dd v\dd z + \sum_{i=1}^{d-1} \dd x_i^2 \right)\ .
\end{equation}
By the time-translation symmetry of the eternal geometry, we may choose a symmetric anchoring convention as illustrated in Figure \ref{fig:CV-latetime}. With our orientation of the Schwarzschild time coordinate, we write $-t_L= t_R = t_B \geqslant 0$ by symmetry. Owing to the translational symmetry along the spatial coordinates $x^i$, the induced metric on this slice can be parameterized by a parameter $\lambda$, such that $v = v(\lambda)$ and $z = z(\lambda)$. The volume density of the co-dimension 1 surface $V_{d-1}^\text{CV} = V_\text{CV}/\mathcal{V}_{d-1}$ is then given by the integration of Lagrangian $L$:
\begin{equation}\label{eq:CV_d-1}
    \begin{aligned}
        V_{d-1}^\text{CV} & = \int \dd\lambda \,L(v, \dot{v}, z, \dot{z}) \ ,\\
        L_\text{CV}(v, \dot{v}, z, \dot{z}) &= \frac{1}{z^{d}} \sqrt{-f(z)\e^{-\chi(z)}\dot{v}^2 + 2\e^{-\chi(z)/2}\dot{v}\dot{z}} \ .
    \end{aligned}
\end{equation}
Since the Lagrangian $L_\text{CV}$ has no explicit $v$-dependence, the conserved momentum $E$ conjugate to $v$ is conserved:
\begin{equation} \label{eq:Energy}
    E \equiv -\frac{\partial L_\text{CV}}{\partial \dot{v}} = \frac{1}{z^{d}} \frac{f(z)\e^{-\chi(z)}\dot{v} - \e^{-\chi(z)/2}\dot{z}}{\sqrt{-f(z)\e^{-\chi(z)}\dot{v}^2 + 2\e^{-\chi(z)/2}\dot{v}\dot{z}}} \ .
\end{equation}

At the turning point $M$ of the maximal slice inside the horizon, shown in Fig.~\ref{fig:CV-latetime}, we have $\dot{z}|_M=0$. Evaluating the conserved quantity~\eqref{eq:Energy} at this point reduces to a local geometric turning-point function $W_\text{CV}(z_M)$
\begin{equation}
   E_{M} = \frac{\sqrt{-f(z_M)\e^{-\chi(z_M)}}}{z_M^{d}}\equiv W_\text{CV}(z_M)\ .
\end{equation}
Here we define the turning-point function
\begin{equation}\label{eq:WCV_z}
    W_\text{CV}(z) \equiv \frac{\sqrt{-f(z)\e^{-\chi(z)}}}{z^{d}}\ .
\end{equation}

The full trajectory may be integrated to obtain the boundary time $t_B$ as a function of $z_M$, but the growth rate follows directly from the Hamilton-Jacobi relation. Varying the on-shell volume~\eqref{eq:CV_d-1} with respect to the symmetric boundary time $t_B$ yields the conjugate momentum $E$.  Thus, the complexity growth rate is given by:
\begin{equation}
\frac{\dd C_V}{\dd t_B} = \frac{\dd V_{d-1}^\text{CV}}{\dd t_B} = 2W_\text{CV}(z_M)\ .
\end{equation}

The standard late-time CV branch requires a finite interior critical point that controls the limiting turning point of the maximal slice. In other words, on the late-time branch of a static eternal black hole, the boundary time becomes large as the turning point approaches a finite interior radius $z_M\to z_{M,c}, t_B\to\infty$. We denote the corresponding limiting interior surface by a critical extremal surface $\mathcal{A}_c^\text{CV}$, as dictated in Fig.~\ref{fig:CV-latetime}. The critical radius is a stationary point of the turning-point function, $W'_\text{CV}(z_{M,c}) = 0$, or equivalently
\begin{equation}
     z_{M,c} \frac{f'(z_{M,c})}{f(z_{M,c})} - z_{M,c} \chi'(z_{M,c}) - 2d = 0 \ .
\end{equation}
The growth rate therefore approaches
\begin{equation}
     \lim_{t_B\to\infty}\frac{\dd C_V}{\dd t_B} = 2W_\text{CV}(z_{M,c})\ ,
\end{equation}
which gives the well-known asymptotically linear CV growth.

Having reviewed this standard late-time mechanism, we now ask what can be inferred from the terminal Kasner geometry alone. In particular, we seek an endpoint condition that guarantees the existence of at least one positive finite-radius maximum of $W_\text{CV}(z)$, providing the critical point required by the standard late-time mechanism. At the event horizon, $f(z_h)=0$ gives $W_\text{CV}(z_h)=0$, while inside the horizon $z_h < z < \infty$, the blackening factor is negative $f(z) < 0$, ensuring that $W_\text{CV}(z) > 0$. If $W_\text{CV}(z)$ also vanishes at the singularity, i.e., $\lim_{z \to \infty} W_\text{CV}(z) = 0$,  continuity guarantees\footnote{If several extrema are present, identifying the branch relevant at late times requires the full relation between $t_B$ and $z_M$. The endpoint argument used here guarantees existence, but does not select among multiple competing branches.} that $W_{\rm CV}$ reaches a positive maximum at some finite interior radius $z_{M,c}$. As the turning point $z_M$ approaches such a stationary maximum $z_{M,c}$, the radial integral determining the boundary time develops the standard critical behavior, so $t_B$ can become arbitrarily large.

Near the singularity, utilizing the asymptotic expansion in Eq.~\eqref{eq:assumofchi-f}, we have:
\begin{equation}
    W_\text{CV}(z) \sim \sqrt{f_\infty \e^{-\chi_\infty}}\,z^{a+b+1-d} = \sqrt{f_\infty \e^{-\chi_\infty}}\,z^{-\frac{(d-1)p_s + p_t}{p_s}}\ .
\end{equation}
Since $p_s > 0$, the requirement $\lim_{z\to\infty}W_\text{CV}(z) = 0$ demands a positive exponent:
\begin{equation}\label{eq:CVbound}
    (d-1)p_s + p_t > 0\ .
\end{equation}
Equation~\eqref{eq:CVbound} therefore provides an endpoint criterion that guarantees the existence of at least one finite-radius CV critical point. If the inequality is violated, the terminal scaling alone no longer determines whether such a critical point exists, and the full radial profile of $W_\text{CV}(z)$ must be examined. When several extrema are present, the full relation between $t_B$ and $z_M$ is also needed to identify the branch relevant at late times.

\subsection{Hartman-Maldacena Entropy Growth}\label{subsec:HM}

\begin{figure}[htbp]
 \begin{center}
   \includegraphics[width=1\textwidth]{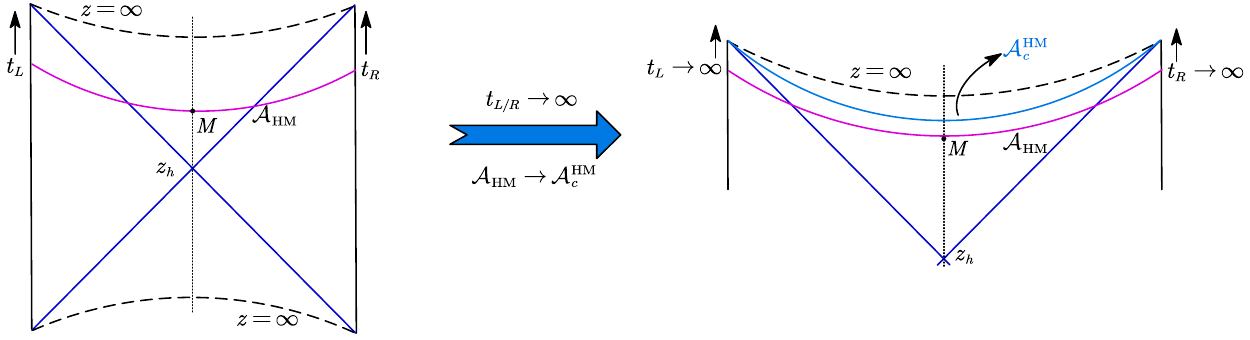}
 \end{center}
\caption{A connected Hartman-Maldacena extremal surface joining the two asymptotic boundaries at symmetric boundary times. The surface is homogeneous along $x^2,\ldots,x^{d-1}$ and reaches its maximum radial depth at $z=z_m$. Along the standard connected late-time branch, $z_m$ approaches the finite interior extremum $z_\text{ext}$ that controls the asymptotic area growth.} \label{fig:HM-TEE-latetime}
\end{figure}

The Hartman-Maldacena (HM) surface provides a second interior-sensitive holographic probe~\cite{Hartman:2013qma, Li:2022cvm}. For a two-sided thermal state, it computes the entanglement entropy of a pair of half-space regions. The gravity dual of HM entanglement entropy is the area of an extremal co-dimension 2 surface $\Gamma$ extending through the interior in a two-sided planar black hole, as illustrated in Fig.~\ref{fig:HM-TEE-latetime}. The entanglement entropy is defined as:
\begin{equation}
   S_\text{HM} = \frac{\text{Area}(\Gamma)}{4 G_\text{N}}\ .
\end{equation}
We find the geometric diagnostic for the HM entanglement entropy closely parallels the logic of the CV complexity. The difference is that the HM surface has one fewer transverse spatial direction than the maximal volume slice. This changes the radial weight in the area functional and, as we will see, the corresponding Kasner criterion.

For a planar entangling surface at $x^1=0$, the HM surface is homogeneous along $x^2,\ldots,x^{d-1}$. Using the same symmetric boundary-time convention as in the CV analysis and parameterizing one half of the surface in Fig.~\ref{fig:HM-TEE-latetime} by $t$, the area density $\mathcal{A}_\text{HM}=\text{Area}(\Gamma)/\mathcal{V}_{d-2}$ is given by:
\begin{equation}
    \begin{aligned}
        \mathcal{A}_\text{HM} &= 2 \int_{0}^{t_B} \dd t \,L(z,\dot{z})\ ,\\
         L_\text{HM}(z,\dot{z}) &= \frac{1}{z^{d-1}}\sqrt{-f(z)\e^{-\chi(z)}+\frac{\dot{z}^2}{f(z)}}\ .
    \end{aligned}
\end{equation}
Because the integrand has no explicit $t$ dependence, the corresponding Hamiltonian $H= \dot{z}\frac{\partial L}{\partial\dot{z}} - L_\text{HM}$ is conserved. By symmetry, there exists a turning point $m$ where $z=z_m$ reaches a local maximum inside the horizon and $t_m=0$ (see Fig.~\ref{fig:HM-TEE-latetime}). At this point, $\dot{z}|_m=0$.
Evaluating $H$ at the point $m$ yields the conserved value:
\begin{equation}\label{eq:HatM}
    \begin{aligned}
        H|_{z=z_m} &= -G_\text{HM}(z_m)\, \\
        G_\text{HM}(z) &\equiv \frac{\sqrt{-f(z)\e^{-\chi(z)}}}{z^{d-1}} \ .
    \end{aligned}
\end{equation}
Here, we introduce a turning-point function $G_\text{HM}(z)$.

The Hamilton-Jacobi relation for the two symmetric halves gives
\begin{equation}\label{eq:growthrate}
    \frac{\dd \mathcal{A}_\text{HM}}{\dd t_B} = -2H = 2G_\text{HM}(z_m)\ .
\end{equation}
The apparent singularity at the horizon in the Schwarzschild-time integral for $t_B(z_m)$ is a coordinate artifact and disappears in regular Eddington-Finkelstein coordinates. The large-$t_B$ limit instead occurs when the turning point approaches an interior extremum of $G_\text{HM}(z)$.
Repeating the CV turning-point argument, the standard connected late-time branch approaches a finite-radius extremum $z_\text{ext}$ of the function $G_\text{HM}(z)$ as $t_B\to\infty$. Its location satisfies $G_\text{HM}'(z_\text{ext})=0$,
\begin{equation}
    2(d-1)f(z_\text{ext}) - z_\text{ext}f'(z_\text{ext}) + z_\text{ext}f(z_\text{ext})\chi'(z_\text{ext}) = 0 \ .
\end{equation}
This is the geometric origin of the familiar linear late-time HM entropy growth.

As in the CV analysis, $G_\text{HM}(z=z_h)=0$ at the horizon. We therefore ask for an endpoint condition that guarantees at least one finite-radius positive maximum
\begin{equation}\label{eq:HMSing}
    \lim_{z \to \infty} G_\text{HM}(z) = 0\ .
\end{equation}
By substituting the near-singularity asymptotic scaling relations~\eqref{eq:assumofchi-f} into $G_\text{HM}(z)$, we can express it as
\begin{equation}
    G_\text{HM}(z) \sim \sqrt{f_\infty \e^{-\chi_\infty}} \,z^{-\frac{p_t+(d-2)p_s}{p_s}}\ .
\end{equation}
Since $p_s>0$, the limit~\eqref{eq:HMSing} requires the combined exponent to be positive:
\begin{equation}\label{eq:HMbound}
    p_t + (d-2)p_s > 0\ .
\end{equation}
Equation~\eqref{eq:HMbound} therefore gives an endpoint criterion that guarantees the existence of at least one finite-radius HM critical point associated with the standard connected late-time mechanism. If the inequality is violated, the terminal scaling alone no longer guarantees such a critical point, and the full profile of $G_\text{HM}(z)$ must be examined.

The CV and HM inequalities therefore arise from the same endpoint mechanism: in each case, decay of the turning-point function at the singularity guarantees at least one finite-radius interior maximum. Such a critical point is the geometric ingredient underlying the standard late-time branch. The different Kasner combinations reflect the different numbers of transverse directions wrapped by the two surfaces. We next turn to CA, for which the singular endpoint enters the construction directly rather than through the existence of an intermediate extremum.

\subsection{Bulk Action Contribution to Complexity = Action}\label{subsec:CA}

The CV and HM analyses above are controlled by critical extremal surfaces. Complexity=Action (CA) probes the terminal interior more directly, because the Wheeler-DeWitt (WdW) patch of CA conjecture can extend to the space-like singularity.

\begin{figure}[htbp]
\centering
\includegraphics[width=0.9\linewidth]{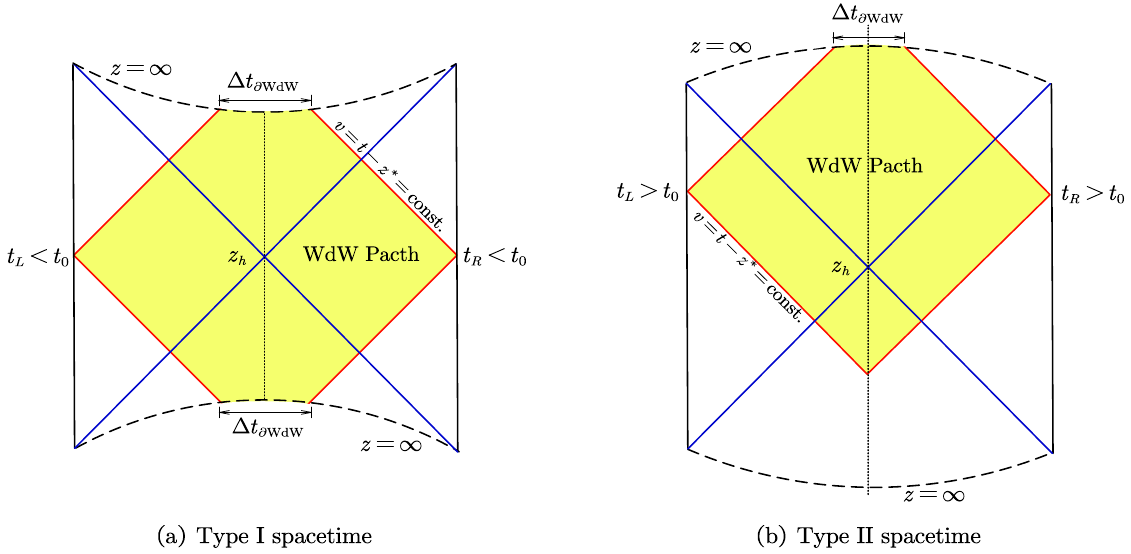}
\caption{Representative WdW patches for Type-I and Type-II black-hole interiors. The two classes differ in the causal shape of the space-like terminal singularity. As a result, the null boundaries of the WdW patch meet the terminal region differently. The detailed causal classification and the associated critical times are reviewed in Appendix~\ref{app:CABulk}.} \label{fig:Type-I-II-CA}
\end{figure}

The Complexity=Action proposal~\cite{Brown:2015bva,Brown:2015lvg} identifies the quantum complexity of a boundary state with the evaluated on-shell gravitational action within the corresponding Wheeler-DeWitt patch:
\begin{equation}
    C_A = \frac{I_\text{WdW}}{\pi \hbar} \ .
\end{equation}
The WdW patch is defined as the causal domain of dependence of any bulk Cauchy slice anchored at the specific boundary time, as depicted in Fig.~\ref{fig:Type-I-II-CA}. As pointed out by Ref.~\cite{Lehner:2016vdi}, due to the presence of boundaries and intersections, the total action is constructed as:
\begin{equation}\label{eq:IWdW}
    I_\text{WdW} = I_\text{bulk} + I_\text{bdy} + I_\text{joint} + I_\text{ct}\ ,
\end{equation}
where $I_\text{bdy}$ denotes the boundary action (time-like, space-like, or null), $I_\text{joint}$ accounts for the joints connecting these boundaries, and $I_\text{ct}$ is the counterterm required for reparameterization invariance.

The global way in which the WdW patch reaches the terminal region depends on the causal type of the black hole interior, as illustrated in Fig.~\ref{fig:Type-I-II-CA}.
For the class of WdW patches considered here, all null boundary, joint, and counterterm contributions supported at finite $z$ remain finite~\cite{Lehner:2016vdi, Yang:2017amx, Yang:2017czx}. We further require that these terms remain finite independently, i.e., without relying on cancellations among divergent contributions.  The only possible loss of well-definedness comes from the portion of the regulated WdW action that approaches the space-like singularity, as shown in Fig.~\ref{fig:Type-I-II-CA}.  In this work, we therefore isolate the bulk contribution $I_\text{bulk}$, which is the term most directly sensitive to the scaling behavior of the spacetime singularity.

To isolate this critical bulk action contribution, $I_\text{bulk} = \int \dd^{d+1}x \,\sqrt{-g} \, L$, we consider the generic Lagrangian density $16\pi L = R + d(d-1) + L_\text{matter}$. We assume the matter fields preserve the transverse $\text{SO}(d-1)$ homogeneity and isotropy of the planar metric ansatz, which implies that the spatial stress tensor takes the form ${T^i}_j = P(z) {\delta^i}_j$, where $P(z)$ serves as an effective local transverse pressure. Motivated by the homogeneous matter configurations relevant to many standard holographic backgrounds, we further restrict to matter configurations $L_\text{matter}$ with no active transverse gradients or tensor components. For this class of backgrounds, we impose:
\begin{equation}\label{eq:matter_condition}
    \frac{\partial L_\text{matter}}{\partial g^{ij}} = 0\ .
\end{equation}
This condition is satisfied by many standard homogeneous holographic backgrounds, including radial scalar profiles and purely electric Einstein-Maxwell-dilaton configurations. Its detailed scope and several examples are discussed in Appendix~\ref{app:CABulk}.

Under the condition~\eqref{eq:matter_condition}, the transverse Einstein equation and the trace equation allow the on-shell bulk Lagrangian to be written as the exact radial derivative
\begin{equation}\label{eq:bulk_radial_derivative}
    16\pi L\sqrt{-g} = 2\frac{\dd}{\dd z}\left( \e^{-\chi/2}f(z)z^{-d} \right)\ .
\end{equation}
A derivation of Eq.~\eqref{eq:bulk_radial_derivative} is given in Appendix~\ref{app:CABulk}.
Thus, the bulk action evaluated over the WdW patch (factoring out the transverse spatial volume $\V_{d-1}$) simplifies to:
\begin{equation}
    \frac{I_\text{bulk}}{\V_{d-1}} = \frac{1}{8\pi} \iint_\text{WdW} \dd t \, \dd z \left[ \frac{\dd}{\dd z}\left( \e^{-\chi/2}f(z)z^{-d} \right) \right]\ .
\end{equation}
Applying Green's theorem in the $(t,z)$ plane,  the volume integral reduces to boundary terms evaluated at the edges $\partial \text{WdW}$ of the WdW patch, as shown in Fig.~\ref{fig:Type-I-II-CA}. At large $z$, the singularity-side contribution to the bulk action takes the form
\begin{equation}
    \frac{I_\text{bulk}}{\V_{d-1}} = \frac{\Delta t_{\partial\text{WdW}}}{8\pi}\cdot\lim_{z\to \infty}H_\text{CA}(z)\ ,
\end{equation}
where $\Delta t_{\partial\text{WdW}}(z)$ is the coordinate-time width of the regulated WdW patch, and the function $H_\text{CA}(z)$ is defined by
\begin{equation}
    H_\text{CA}(z) \equiv \e^{-\chi(z)/2}f(z)z^{-d}\ .
\end{equation}
If the width $\Delta t_{\partial\text{WdW}} $ approaches a nonzero constant, finiteness of the bulk contribution requires $H_\text{CA}(z)$ itself to remain finite as approaching the singularity $z\to \infty$.  More generally, finiteness of $H_\text{CA}$ is a sufficient condition, which we adopt here as a local criterion on the bulk action density; it does not depend on the detailed shape of the WdW patch near the endpoint.

Using our asymptotic scaling assumptions~\eqref{eq:assumofchi-f} near the singularity, we have
\begin{equation}
    H_\text{CA}(z) \sim-f_\infty \e^{-\chi_\infty/2} \,z^{2a + b + 2 - d}\sim - f_\infty \e^{-\chi_\infty/2} \,z^{-\frac{(d-1)p_s+ p_t-1}{p_s}}\ .
\end{equation}
Because $p_s > 0$, requiring $H_\text{CA}(z)$ to remain finite imposes the constraint:
\begin{equation}\label{eq:CAbound}
    (d-1)p_s + p_t \geqslant 1\ .
\end{equation}
Equation~\eqref{eq:CAbound} is therefore the Kasner criterion associated with the local finiteness of the CA bulk contribution. If $(d-1)p_s+p_t>1$, $H_\text{CA}(z)$ vanishes toward the singularity, while equality gives a finite nonzero limiting value. If $(d-1)p_s+p_t<1$, $H_\text{CA}(z)$ diverges. Since the remaining boundary, joint, and counterterm contributions are finite in the class of WdW patches considered here, this divergence makes the WdW action, and hence the CA complexity, ill-defined. Equation~\eqref{eq:CAbound} therefore gives the Kasner condition required for a finite CA complexity near the space-like singularity.

\subsection{Holographic Thermal \texorpdfstring{$a$}{a}-Function and the Trans-IR Endpoint}

The preceding constructions are non-local functionals of bulk surfaces or spacetime regions.
The thermal $a$-function instead provides a local geometric diagnostic of the holographic radial flow~\cite{Caceres:2022smh,Caceres:2022hei,Caceres:2023zft}. In the trans-IR interpretation, the horizon marks the IR endpoint of the exterior holographic flow, and analytic continuation through the horizon extends this flow into the black hole interior toward the space-like singularity. Within this construction, the thermal $a$-function is interpreted as a measure of the effective degrees of freedom along the analytically continued flow.

In the radial $z$-coordinate used throughout this work, the thermal $a$-function takes the form~\cite{Caceres:2022smh}:
\begin{equation}\label{eq:thermala}
    a_T(z) = \frac{\pi^{d/2}}{\Gamma\left(\frac{d}{2}\right)\ell_P^{d-1}} \e^{-(d-1)\chi(z)/2}\ .
\end{equation}
In the original construction, the radial null energy condition guarantees the monotonicity of $a_T$ along the flow\footnote{In a standard domain-wall geometry parameterized by an energy scale $u$, the null energy condition (NEC) ensures that $a_T(u)$ decreases monotonically along the flow~\cite{Caceres:2022smh}. Here to seamlessly connect this trans-IR flow to our near-singularity analysis, we map the domain-wall metric to our standard radial $z$-coordinate.}. Here we first examine its terminal behavior directly from the Kasner geometry; its relation to the null energy condition will be discussed in Sec.~\ref{sec:EnergyCondition}.

Physically, a sensible trans-IR flow must smoothly terminate at the space-like singularity. At this spacetime singularity, the effective degrees of freedom are expected to remain finite. By substituting our near-singularity Kasner scaling relation into the $a$-function, we determine its asymptotic behavior:
\begin{equation}
    a_T(z) \sim z^{b(d-1)} \sim \e^{-(d-1)\chi_\infty/2} \,z^{(d-1)\frac{p_s - p_t - 1}{p_s}}\ .
\end{equation}
For $a_T(z)$ to remain finite at the singularity ($z \to \infty$), the scaling exponent must be non-positive (noting that $p_s>0$), which directly yields:
\begin{equation}\label{eq:aTbound}
    p_t \geqslant p_s - 1\ .
\end{equation}
More explicitly, the three possible asymptotic behaviors of $a_T(z)$ at the singular terminal boundary are categorized by:
\begin{equation}
    \begin{dcases}
        p_t > p_s - 1\ , \implies a_T(z) \to 0\ ;  \\
        p_t = p_s - 1\ , \implies a_T(z) \to \text{finite constant}\ ; \\
        p_t < p_s - 1\ , \implies a_T(z) \to \infty\ .
     \end{dcases}
\end{equation}
Thus, the bound~\eqref{eq:aTbound} excludes geometries where $a_T(z)$ diverges near the singularity, while accommodating cases where $a_T(z)$ saturates to a finite non-zero constant. In particular, the Schwarzschild Kasner exponents lie on the equality line, serving as a classic example where $a_T(z)$ approaches a non-zero finite value at the singularity. It therefore provides a clear condition for maintaining a well-behaved measure of effective degrees of freedom in the deep interior. If the inequality is violated, $a_T(z)$ diverges as $z \to \infty$, rendering its dual field-theory interpretation ill-defined in the asymptotic interior regime.

\subsection{Summary of Holographic Kasner Bounds}

\begin{table}[t]
  \centering
  \renewcommand{\arraystretch}{1.45}
  \setlength{\tabcolsep}{3.5pt}
  \begin{tabular}{@{}
    >{\centering\arraybackslash}p{0.12\linewidth}
    >{\centering\arraybackslash}p{0.31\linewidth}
    >{\raggedright\arraybackslash}p{0.33\linewidth}
    c
  @{}}
    \toprule
    \textbf{Observable}
    & \textbf{Controlling function}
    & \textbf{Role of the sufficient criterion}
    & \textbf{Kasner bound}
    \\
    \midrule
    CV
    & $\displaystyle W_{\mathrm{CV}}(z)\equiv\frac{\sqrt{-f(z)\,\e^{-\chi(z)}}}{z^d}$
    & Guarantees an interior critical surface $\mathcal{A}^\text{CV}_c$
    & $\displaystyle (d-1)p_s+p_t > 0$
    \\[6pt]

    HM
    & $\displaystyle G_{\mathrm{HM}}(z)\equiv\frac{\sqrt{-f(z)\,\e^{-\chi(z)}}}{z^{d-1}}$
    & Guarantees an interior critical surface $\mathcal{A}^\text{HM}_c$
    & $\displaystyle p_t+(d-2)p_s > 0$
    \\[6pt]

    CA bulk
    & $\displaystyle H_{\mathrm{CA}}(z)\equiv \e^{-\chi(z)/2}f(z)z^{-d}$
    & Finite near-singularity contribution to bulk action
    & $\displaystyle (d-1)p_s+p_t\geqslant 1$
    \\[6pt]

    Thermal $a$-Function
    & $\displaystyle a_T(z)\sim \e^{-(d-1)\chi(z)/2}$
    & Finite terminal value of the trans-IR $a$-function
    & $\displaystyle p_t \geqslant p_s-1$
    \\
    \bottomrule
  \end{tabular}
  \caption{%
    The Kasner bound, an algebraic constraint on the Kasner exponents $(p_t,p_s)$ obtained from four interior-sensitive holographic observables. For each interior-sensitive observable, the near-singularity behavior is governed by a specific metric combination. For CV and HM, the inequalities are sufficient endpoint conditions that guarantee the existence of a critical extremal surface associated with the turning-point function. For the CA bulk contribution and the thermal $a$-function, the inequalities instead control quantities evaluated directly toward the terminal interior.
  }
  \label{tab:holographic-kasner-bounds}
\end{table}

The four Kasner criteria derived in this section naturally separate according to how the corresponding holographic constructions probe the interior. For CV and HM, the terminal Kasner scaling controls the endpoint behavior of the turning-point functions: when these functions vanish at both the horizon and the singularity, an intermediate finite-radius critical point is guaranteed. The CA and thermal $a$-function constructions instead probe the terminal region directly, so their Kasner bounds determine whether the corresponding quantities remain finite as the singularity is approached. The resulting conditions are summarized in Table~\ref{tab:holographic-kasner-bounds}.

These four bounds have been obtained entirely from the holographic constructions and the assumed terminal Kasner geometry, without imposing any classical energy condition. This naturally raises the next question: can familiar bulk energy conditions constrain the Kasner exponents so that these holographic requirements are automatically satisfied? In Sec.~\ref{sec:EnergyCondition}, we show that the null and dominant energy conditions provide precisely such criteria.

\section{Energy Conditions and Holographic Kasner Criteria}\label{sec:EnergyCondition}

In Sec.~\ref{sec:HolographicBound}, we established that the simple Kasner conditions on $(p_t, p_s)$ guarantee the geometric or endpoint properties relevant to the four holographic constructions considered there.  If a model violates these bounds, several possibilities arise: the semi-classical near-singularity description (Kasner/Einstein) may break down before the bounds are reached; the holographic prescription under consideration may not apply; or the bulk geometry may have no consistent boundary dual. 

The analysis of Sec.~\ref{sec:HolographicBound} identifies regions of the Kasner parameter space associated with the geometric and endpoint behavior of the four interior-sensitive holographic constructions. However, in the bottom-up approach, simply writing down an arbitrary effective bulk action does not guarantee that its resulting black hole interior will satisfy these geometric requirements. Therefore, a practical question arises: what physical properties of a bulk gravitational theory can directly guarantee that these Kasner bounds are automatically satisfied?  
We therefore turn to the bulk dynamics and ask how these regions are constrained by the matter supporting the black hole geometry. Classical energy conditions provide a natural link between the two: through the Einstein equations, local constraints on the stress tensor translate directly into inequalities on the terminal Kasner exponents.

We evaluate the null and dominant energy conditions in the Kasner interior and compare the resulting inequalities with the holographic bounds derived in Sec.~\ref{sec:HolographicBound}. We will show that the transverse NEC implies the CV and HM bounds, while the radial NEC reproduces the thermal $a$-function bound. The radial DEC similarly gives the CA bound. In this way, the classical energy conditions provide simple bulk criteria that place the Kasner exponents within the holographically allowed regions selected by the corresponding holographic probes. 

\subsection{Energy-Condition Inequalities in the Kasner Interior}\label{subsec:nullframes}

In this subsection, we will derive the energy-condition constraints directly using the holographic metric functions $f(z)$ and $\chi(z)$ defined in Eq.~\eqref{eq:AdSmetric}. For an independent and shorter geometric proof formulated in terms of the local energy density and principal pressures of a general Kasner universe, we direct the reader to Appendix~\ref{app:KasnerEnergy}.

Inside the event horizon where $f(z) < 0$, the radial coordinate $z$ becomes time-like, while the coordinate $t$ becomes space-like. Consequently, we can construct normalized orthogonal basis vectors (recall $i=1,\ldots,d-1$):
\begin{equation}
    \begin{aligned}
        e_{\hat{0}}^\mu &= \frac{1}{\sqrt{-g_{zz}}}(\partial_z)^\mu = \left(0, z\sqrt{-f}, \dots\right)\ ;\\
        e_{\hat{1}}^\mu &= \frac{1}{\sqrt{g_{tt}}}(\partial_t)^\mu = \left(z\e^{\chi/2}/\sqrt{-f}, 0, \dots\right)\ ;\\
        e_{\hat{i}}^\mu &= \frac{1}{\sqrt{g_{x^ix^i}}}(\partial_{x^i})^\mu = (0, \dots, z, \dots)\ .
    \end{aligned}
\end{equation}
Inside the horizon, the $e_{\hat{0}}^\mu$ frame vector is time-like, while $e_{\hat{1}}^\mu$ and $e_{\hat{i}}^\mu$ are space-like. They satisfy
\begin{equation}
    e_{\hat{0}}\cdot e_{\hat{0}}=-1\ , \qquad e_{\hat{1}}\cdot e_{\hat{1}} = e_{\hat{i}}\cdot e_{\hat{i}} =1\ .
\end{equation}
We can therefore form two useful classes of null vectors in the $z-t$ plane (spanned by $z$ and $t$ ) and $z-x^i$ plane (spanned by the $x^i$, $i=1,\cdots,d-1$):
\begin{equation}
    \begin{aligned}
        l_{(z,t)}^\mu &= \frac{1}{\sqrt{2}}(e_{\hat{0}}^\mu + e_{\hat{1}}^\mu)\ ,\qquad k^\mu_{(z,t)} = \frac{1}{\sqrt{2}}(e_{\hat{0}}^\mu - e_{\hat{1}}^\mu) \ ;\\
        l^\mu_{(z,x^i)} &= \frac{1}{\sqrt{2}}(e_{\hat{0}}^\mu + e_{\hat{i}}^\mu)\ ,\qquad   k^\mu_{(z,x^i)} = \frac{1}{\sqrt{2}}(e_{\hat{0}}^\mu - e_{\hat{i}}^\mu)\ .
    \end{aligned}
\end{equation}
It is straightforward to verify that these satisfy the standard null conditions $l_\mu l^\mu = 0$, $k_\mu k^\mu = 0$, and $l_\mu k^\mu = -1$.

From the Einstein equations $G_{\mu\nu} + \Lambda g_{\mu\nu} = \frac{1}{2}T_{\mu\nu}$, the null energy condition (NEC) evaluated within the $z-t$ plane, $T_{\mu\nu} l^\mu_{(z,t)} l^\nu_{(z,t)} \geqslant 0$, imposes a direct bound. Substituting the metric functions yields (where the prime denotes the derivative with respect to $z$):
\begin{equation}
   \frac{1}{2} T_{\mu\nu} l_{(z,t)}^\mu l_{(z,t)}^\nu = -\frac{d-1}{2} z f(z)\chi^{\prime}(z) \geqslant 0\ .
\end{equation}
NEC requires $\chi'(z) \geqslant 0$ (because inside horizon $f(z)<0$). Given $\chi(z) \sim -2b\ln z+\chi_\infty$ near the space-like singularity, this restricts $b \leqslant 0$. Since $1+p_t-p_s=-b p_s$ (see Eq.\eqref{eq:definitionp_sp_t}) and $p_s>0$, the Kasner inequality is
\begin{equation}\label{eq:NECzt}
    p_s - p_t \leqslant 1\ .
\end{equation}
As explicitly shown in Appendix~\ref{app:KasnerEnergy}, this inequality is equivalent to the longitudinal null energy condition $\rho + P_t \geqslant 0$ in the local orthonormal frame.

The NEC statement $T_{\mu\nu} l^\mu_{(z,x^i)} l^\nu_{(z,x^i)} \geqslant 0$ applied to the transverse $z-x^i$ plane yields (where the prime denotes the derivative with respect to $z$):
\begin{equation}
    \e^{\chi/2} z^{d-1} \frac{\dd}{\dd z} \left[ \e^{-\chi/2} z^{1-d} (f' - f\chi') \right] - \frac{d-1}{z} f \chi' \geqslant 0\ .
\end{equation}
Substituting the asymptotic scaling behaviors, we obtain $b^2 + (3a+2)b - (a+1)(d - 2a - 2) \leqslant 0$. Translating this into the Kasner exponents we have
\begin{equation}\label{eq:NECzx}
    p_t^2 - (p_s + 1)p_t - (d-2)p_s \leqslant 0\ .
\end{equation}
This inequality restricts $p_t$ to the interval $p_t\in[p_{t,-}, p_{t,+}]$, where
\begin{equation}
    p_{t,\pm} = \frac{1}{2}\left[ p_s+1 \pm \sqrt{(p_s+1)^2+4(d-2)p_s} \right]\ .
\end{equation}
This inequality corresponds geometrically to the requirement that the effective local energy density dominates the transverse null-energy combination ($\rho + P_s \geqslant 0$), as derived in the purely cosmological Kasner framework in Appendix~\ref{app:KasnerEnergy}.

We will also need a radial DEC condition. The DEC $T_{\mu\nu}l^\mu k^\nu\geqslant 0$ within the $z-t$ plane yields
\begin{equation}
        \frac{1}{2} T_{\mu\nu} l^\mu k^\nu = (d-1) z^{d+1} \e^{\chi/2} \left( \frac{f}{z^d} \e^{-\chi/2} \right)^{\prime} + d(d-1)\ .
\end{equation}
As $z\to \infty$, the leading term on the left-hand side scales as $-(2a+2-d+b) z^{2a+2}$. Because $2a+2 > 0$ (required for $a > -1$), the left-hand side must remain non-negative, the coefficient of the leading power must be non-positive; otherwise the expression would run to $-\infty$. Thus, $2(a+1)-d+b \leqslant 0$, which translates to the Kasner exponents as:
\begin{equation}\label{eq:DECzt}
    (d-1)p_s+p_t \geqslant 1\ .
\end{equation}
This is the Kasner inequality implied by the DEC $\rho-P_t\geqslant0$; see Appendix~\ref{app:KasnerEnergy}.

\subsubsection{NEC protection of the CV and HM bounds}

We now show that the transverse NEC is sufficient for the CV and HM bounds. The CV consistency condition derived in Sec.~\ref{subsec:CV} is
\begin{equation}
    (d-1)p_s+p_t > 0 .
\end{equation}
Using the null energy condition in the $z-x^i$ plane as given by Eq.~\eqref{eq:NECzx}, we can verify that the NEC is sufficient to prove the inequality~\eqref{eq:CVbound}:
\begin{equation}
    \begin{aligned}
        p_t + (d-1)p_s &\geqslant \frac{p_s + 1 - \sqrt{(p_s + 1)^2 + 4(d-2)p_s}}{2} + (d-1)p_s\\
        &\geqslant \frac{1}{2} \left[ 1 + (2d-1)p_s - \sqrt{(p_s + 1)^2 + 4(d-2)p_s} \right]\\
        & \geqslant \frac{1}{2} \left[ 1 + (2d-1)p_s - \sqrt{(1 + (2d-1)p_s)^2 - (4p_s + 4d(d-1)p_s^2)} \right] \\
        &> 0\ .
    \end{aligned}
\end{equation}
For any spatial dimension $d \geqslant 3$, the term $4p_s + 4d(d-1)p_s^2$ is manifestly positive, ensuring that the square root is always strictly less than the leading term $1 + (2d-1)p_s$. Thus the transverse NEC does more than guarantee non-divergence of $W_\text{CV}(z)$. It places the Kasner exponents strictly inside the CV-allowed region.

\begin{figure}[htbp]
\centering
\begin{subfigure}[t]{0.47\textwidth}
\centering
\includegraphics[width=\linewidth]{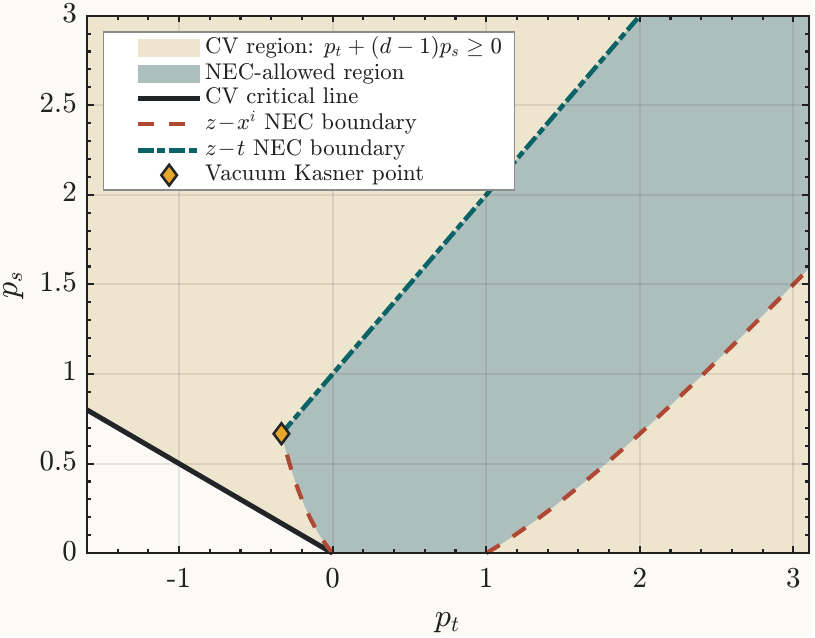}
\caption{$d=3$.}
\end{subfigure}
\qquad
\begin{subfigure}[t]{0.47\textwidth}
\centering
\includegraphics[width=\linewidth]{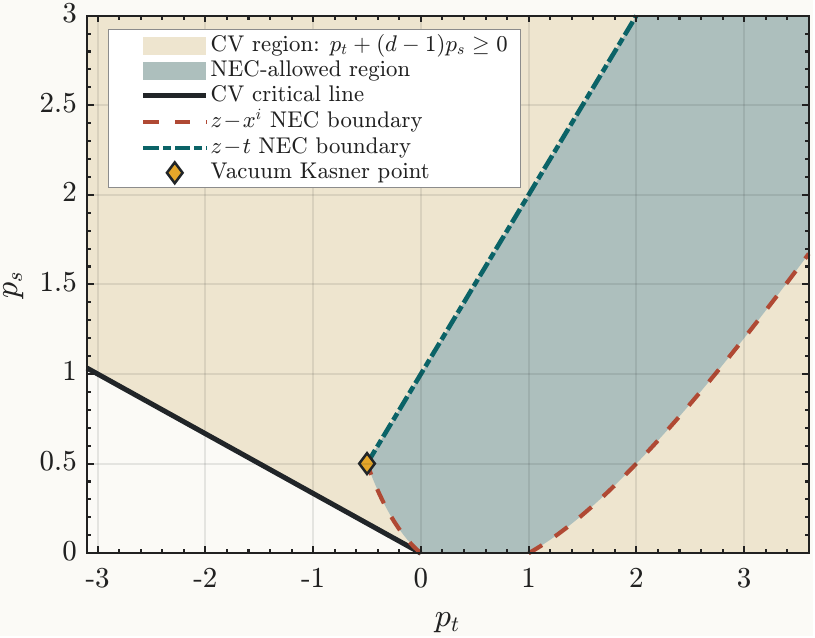}
\caption{$d=4$.}
\end{subfigure}
\caption{Kasner-exponent parameter space for Complexity=Volume (CV) in (a) $d=3$ and (b) $d=4$. The sand-colored half-plane obeys $(d-1)p_s+p_t>0$, which makes $W_\text{CV}(z)$ vanish at the singularity and guarantees an interior maximum governing late-time CV growth. The blue-gray domain is allowed jointly by the transverse and radial NECs; their boundaries are shown by the rust dashed curve and teal dash-dotted line. The gold diamond is the vacuum point $(p_t,p_s)=((2-d)/d,2/d)$. The separation between the NEC domain and the charcoal CV critical line displays the buffer by which the NEC enforces the CV consistency criterion.
} \label{fig:KasnerCV}
\end{figure}

\begin{figure}[htbp]
\centering
\begin{subfigure}[t]{0.47\textwidth}
\centering
\includegraphics[width=\linewidth]{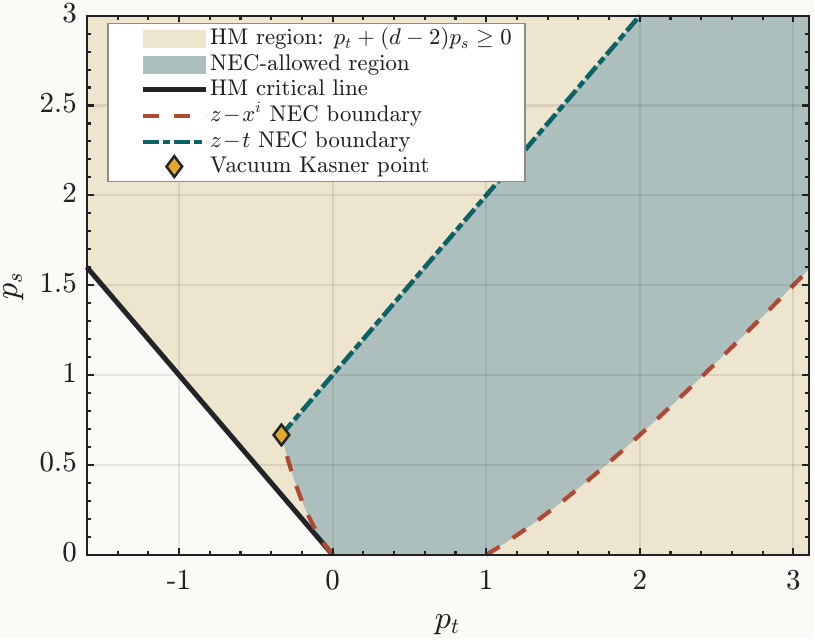}
\caption{$d=3$.}
\end{subfigure}
\qquad
\begin{subfigure}[t]{0.47\textwidth}
\centering
\includegraphics[width=\linewidth]{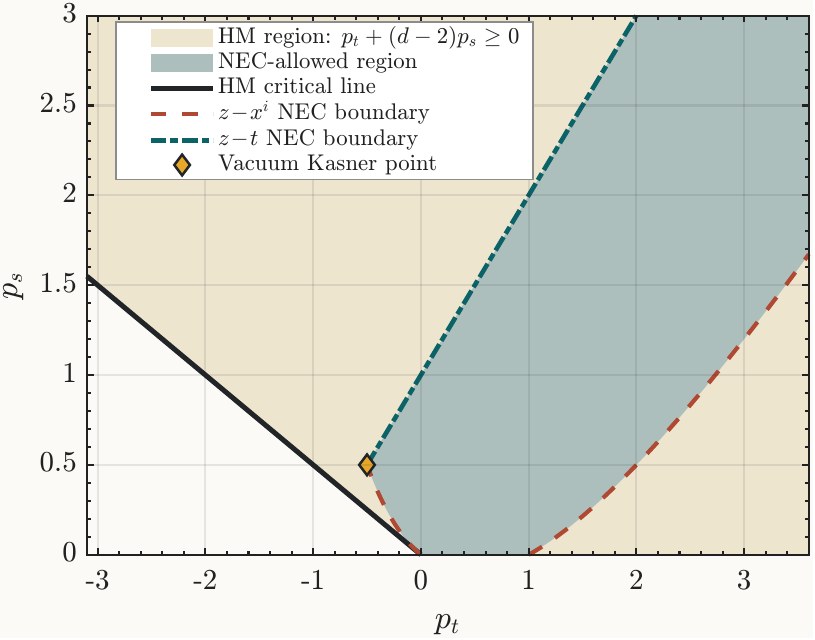}
\caption{$d=4$.}
\end{subfigure}
\caption{Kasner-exponent parameter space for Hartman-Maldacena (HM) entropy in (a) $d=3$ and (b) $d=4$, with $p_t$ on the horizontal axis and $p_s$ on the vertical axis. The sand-colored half-plane satisfies $p_t+(d-2)p_s>0$, for which $G_\text{HM}(z)$ vanishes at the singularity. The blue-gray domain is allowed jointly by the transverse $z-x^i$ and radial $z-t$ null energy conditions (NECs), whose boundaries are the rust dashed curve and teal dash-dotted line. The gold diamond marks the vacuum Kasner point $(p_t,p_s)=((2-d)/d,2/d)$. The NEC domain lies inside the HM-allowed half-plane, showing that the NEC protects the interior critical surface controlling late-time growth. } \label{fig:KasnerHM}
\end{figure}

Similarly, the HM consistency condition derived in Sec.~\ref{subsec:HM} is
\begin{equation}
    p_t+(d-2)p_s > 0 .
\end{equation}
We now verify if the null energy condition (NEC) can protect this central inequality.
Using the lower bound of NEC in the $z-x^i$ plane, we evaluate our target combination:
\begin{equation}
   \begin{aligned}
       p_t + (d-2)p_s &\geqslant \frac{1}{2} \left[ 1+(2d-3)p_s - \sqrt{(p_s+1)^2+4(d-2)p_s} \right]\\
       &= \frac{1}{2} \left[ 1+(2d-3)p_s - \sqrt{\left[1+(2d-3)p_s\right]^2 - 4(d-1)(d-2)p_s^2} \right]\\
       &>0\ .
    \end{aligned}
\end{equation}
For any spatial dimension $d \geqslant 3$, the term $4(d-1)(d-2)p_s^2$ is manifestly positive, ensuring that the square root is always less than the leading term. Thus, $p_t + (d-2)p_s > 0$ follows.
Furthermore, evaluating the NEC in the radial $z-t$ plane yields a distinct lower bound~\eqref{eq:NECzt}. The radial and transverse NEC constraints together define the physically allowed Kasner parameter space (visually summarized in Fig.~\ref{fig:KasnerHM}).

We find that a perturbative or mild violation of the NEC near the singularity will only slightly shift the physical boundaries downward, without crossing the critical line. This ``buffer zone'' mathematically ensures that the critical surfaces $\mathcal{A}^\text{CV}_c$ and $\mathcal{A}^\text{HM}_c$ survive. Consequently, the late-time $C_V$ and $S_\text{HM}$ growth rates remain well-defined, even when classical energy conditions are mildly violated in the deep quantum regime.

\subsubsection{Marginal protection of CA and the thermal \texorpdfstring{$a$}{a}-function}

\begin{figure}[htbp]
\centering
\begin{subfigure}[t]{0.47\textwidth}
\centering
\includegraphics[width=\linewidth]{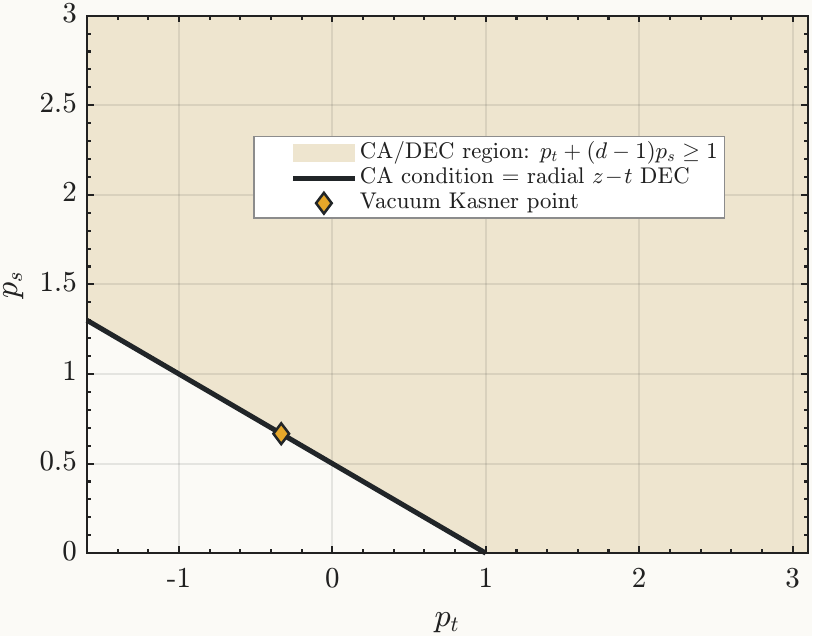}
\caption{$d=3$.}
\end{subfigure}
\qquad
\begin{subfigure}[t]{0.47\textwidth}
\centering
\includegraphics[width=\linewidth]{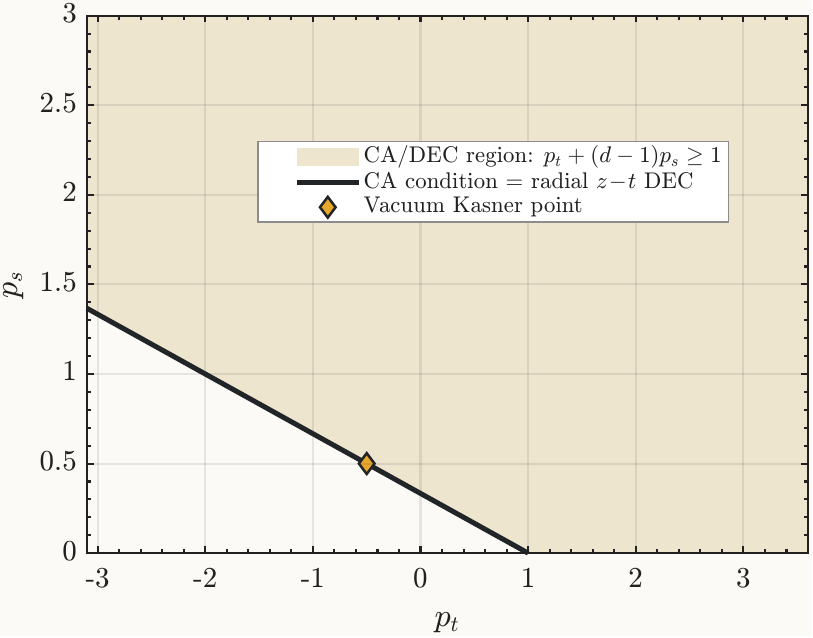}
\caption{$d=4$.}
\end{subfigure}
\caption{Kasner-exponent parameter space for the bulk contribution to Complexity=Action (CA) in (a) $d=3$ and (b) $d=4$. The sand-colored half-plane satisfies $(d-1)p_s+p_t\geqslant 1$, ensuring that $H_\text{CA}(z)$ remains finite at the singularity. The charcoal boundary is simultaneously the CA finiteness threshold and the radial dominant energy condition (DEC), so it is deliberately represented by a single line rather than two superposed curves. The gold diamond marks the vacuum point $((2-d)/d,2/d)$ on this common CA/DEC boundary. The AdS-Schwarzschild point lies on this common boundary.} \label{fig:KasnerCA}
\end{figure}

The situation is different for the CA bulk contribution and the thermal $a$-function. The CA bulk-action bound derived in Sec.~\ref{subsec:CA} is
\begin{equation}
   (d-1)p_s+p_t\geqslant 1\ .
\end{equation}
This coincides with the radial DEC condition~\eqref{eq:DECzt}. Thus the DEC is sufficient for the CA bulk-action bound, but only marginally. There is no finite buffer between the DEC boundary and the CA consistency boundary, as depicted in Fig.~\ref{fig:KasnerCA}.

\begin{figure}[htbp]
\centering
\begin{subfigure}[t]{0.47\textwidth}
\centering
\includegraphics[width=\linewidth]{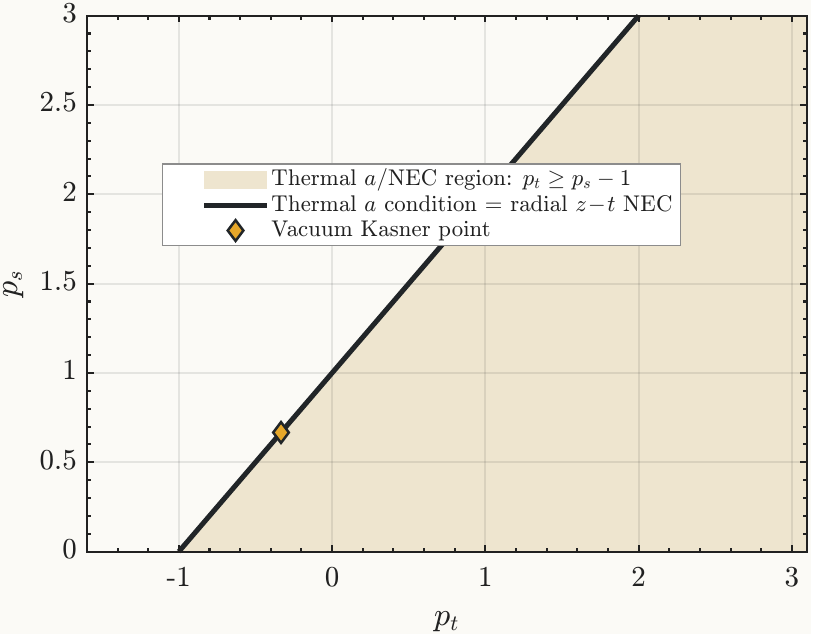}
\caption{$d=3$.}
\end{subfigure}
\qquad
\begin{subfigure}[t]{0.47\textwidth}
\centering
\includegraphics[width=\linewidth]{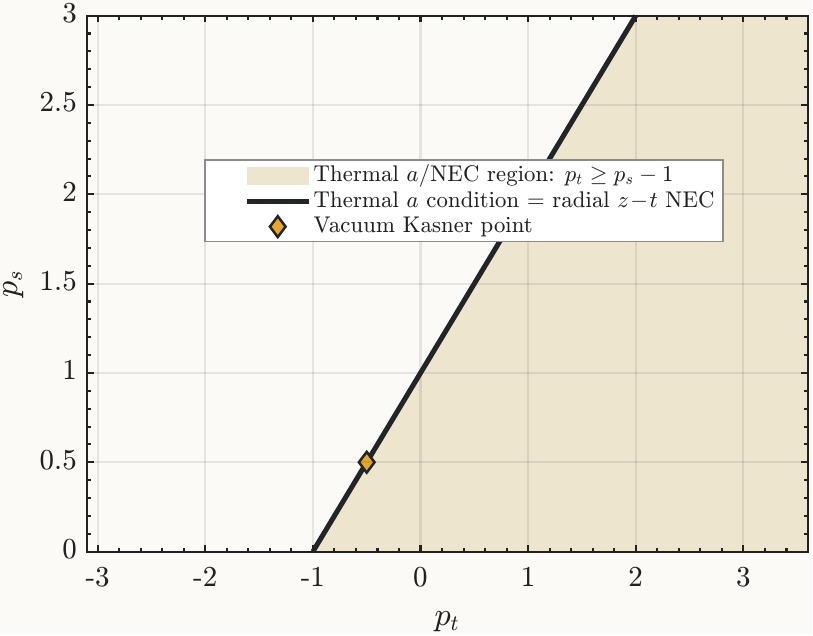}
\caption{$d=4$.}
\end{subfigure}
\caption{ Kasner-exponent parameter space for the thermal $a$-function in (a) $d=3$ and (b) $d=4$. The sand-colored region satisfies $p_t\geq p_s-1$, for which $a_T(z)$ is finite or tends to zero toward the Kasner singularity. Its charcoal boundary is simultaneously the thermal-$a$ regularity threshold and the radial $z-t$ NEC boundary. The gold diamond denotes the vacuum point $(p_t,p_s)=((2-d)/d,2/d)$ on the common boundary. Hence the radial NEC is both necessary and sufficient for a non-divergent terminal trans-IR flow within the assumed scaling regime. } \label{fig:KasnerAT}
\end{figure}

The thermal $a$-function gives another marginal case. Its non-divergence condition is
\begin{equation}
p_t\geqslant p_s-1\ .
\end{equation}
This coincides with the radial NEC condition~\eqref{eq:NECzt} $p_s - p_t \leqslant 1$. Hence, the radial NEC is both necessary and sufficient for a non-divergent terminal trans-IR flow, ensuring the physically sound conclusion that the effective degrees of freedom smoothly extinguish at the Kasner singularity. However, much like the CA bound, this protection is entirely marginal, as shown in Fig.~\ref{fig:KasnerAT}.

This distinction is conceptually important. The CV and HM observables are controlled by extremal geometric objects that settle at intermediate critical radii inside the horizon. Their Kasner bounds are therefore weaker than the full transverse NEC constraint, leaving a finite buffer in the Kasner parameter space. By contrast, the constraints derived from the CA bulk action and the thermal $a$-function provide no such buffer: their required parameter spaces exactly saturate the respective energy condition boundaries.

For example, in the classical vacuum limit (e.g., a neutral Schwarzschild-AdS singularity), the Kasner exponents satisfy the strict equality $(d-1)p_s + p_t = 1$,  placing the spacetime precisely on the marginal edge of this condition. If classical general relativity is treated as an effective field theory, quantum gravity fluctuations or higher-derivative corrections near the singularity are expected to mildly violate the classical energy conditions. The absence of a parameter-space margin for the CA and thermal $a$-function bounds implies that their energy-condition constraints offer no additional robustness beyond the endpoint bound itself. However, this does not necessarily imply that generic quantum or higher-derivative corrections must cross the common threshold. The direction and magnitude of such corrections must be determined within a concrete effective theory.

\subsection{Summary of Sufficient Criteria}

The role of energy conditions is summarized in Table~\ref{tab:energy-sufficient-criteria}. The table should be read in one direction: the energy conditions are sufficient criteria for the holographic Kasner bounds. They are not claimed to be necessary conditions for the existence of a holographic dual.

\begin{table}[h]
  \centering
  \renewcommand{\arraystretch}{1.2}
  \setlength{\tabcolsep}{4.5pt}
  \begin{tabular}{@{} c c l c @{}}
    \toprule
    \textbf{Holographic Bound}
    & \textbf{Energy Condition}
    & \textbf{Role of the Condition}
    & \textbf{Buffer Zone}
    \\
    \midrule

    $(d-1)p_s + p_t > 0$
    & Transverse NEC
    & Protects CV critical surface
    & Yes (Finite)
    \\[3pt]

    $p_t + (d-2)p_s > 0$
    & Transverse NEC
    & Protects HM late-time growth
    & Yes (Finite)
    \\[3pt]

    $(d-1)p_s + p_t \geqslant 1$
    & Radial DEC
    & Ensures finiteness of the CA
    & None (Marginal)
    \\[3pt]

    $p_t \geqslant p_s - 1$
    & Radial NEC
    & Ensures finite thermal $a$-endpoint
    & None (Marginal)
    \\
    \bottomrule
  \end{tabular}
  \caption{%
   Energy conditions as sufficient criteria for the holographic Kasner bounds.
   They are not assumed in deriving the bounds in Table~\ref{tab:holographic-kasner-bounds}; rather, they provide simple bulk diagnostics ensuring that the bounds are satisfied. The presence of a finite buffer zone indicates that the corresponding holographic observable remains robust even if classical energy conditions are mildly violated by perturbative quantum corrections near the singularity.
  }
  \label{tab:energy-sufficient-criteria}
\end{table}

\section{Summary and Discussion}\label{sec:discussion}

In this work, we explored how interior-sensitive holographic observables constrain the Kasner exponents of planar asymptotically AdS black holes. For the isotropic Kasner geometry considered here, the four resulting conditions are summarized in Table~\ref{tab:holographic-kasner-bounds}. For CV and HM, the corresponding inequalities make the turning-point functions vanish at the singularity and thereby guarantee a finite-radius interior critical surface underlying the standard late-time growth. For CA, the Kasner bound ensures that the WdW action remains finite as the singularity is approached. Similarly, the thermal $a$-function bound prevents a divergent endpoint of the trans-IR flow. We then showed that classical energy conditions provide simple bulk criteria for these four Kasner bounds, as summarized in Table~\ref{tab:energy-sufficient-criteria}. Thus, our results suggest that the near-singularity Kasner scaling itself carries nontrivial information about holographic consistency, as probed by these interior-sensitive constructions. In particular, these constructions constrain the holographic allowed Kasner parameter space, while classical energy conditions provide a simple dynamical mechanism that drives the interior geometry into the corresponding regions.

Our energy-condition analysis uncovers a dichotomy in how classical bulk dynamics protect these observables. The transverse NEC implies the CV bound and the HM bound. This creates a finite ``buffer zone'' in the Kasner parameter space: the NEC can be mildly violated while the CV and HM probes remain safely well-defined. By contrast, the bounds for the CA action and the thermal $a$-function exactly coincide with the radial DEC and radial NEC boundaries, respectively, leaving zero buffer. However, we note a necessary caveat: the existence of the CV and HM buffer zones demonstrates an algebraic tolerance in the parameter space, but it does not serve as a general proof of stability against arbitrary quantum or higher-derivative corrections.

Finally, the physical meaning of these bounds depends on how each construction probes the interior. For CV and HM, violation of the endpoint condition means that the terminal Kasner scaling alone no longer determines whether the relevant finite-radius critical surface exists, and the full interior profile becomes important. For CA and the thermal $a$-function, the corresponding inequalities directly control the terminal behavior, and their violation produces a divergence as the singularity is approached. Within the static, planar, and transversely isotropic Kasner geometries considered here, these results make the terminal Kasner scaling a useful complementary diagnostic of the semi-classical interior behavior of bottom-up holographic models.

The structure uncovered above suggests several natural extensions. One immediate direction is to broaden the class of interior-sensitive observables beyond the four constructions studied here. Complexity=Anything provides a general family of codimension-one and codimension-zero observables~\cite{Belin:2021bga, Belin:2022xmt}, in which the functional selecting the extremal surface can be chosen independently of the functional evaluated on that surface. In the examples studied in Ref.~\cite{Jorstad:2023kmq}, when the usual finite-radius late-time extremum disappears, the maximizing surfaces can move toward the boundary of the allowed phase space and approach the space-like singularity. Regulated observables at the boundary of this phase space and families of constant-mean-curvature slices therefore provide complementary ways to probe the geometry arbitrarily close to the singularity. It would be interesting to classify these constructions directly in terms of their Kasner scaling. In particular, when the CV endpoint criterion of Sec.~\ref{sec:HolographicBound} is violated, one should first determine whether the full control function still admits a relevant finite-radius extremum. If not, the system may instead enter a different branch in which the dominant surface is driven toward the singular region. The branch exchanges found for generalized volume observables provide a useful precedent for such a phase structure~\cite{Wang:2023eep,Jiang:2023jti}.

The same viewpoint naturally extends beyond transverse isotropy. A general Kasner interior takes the form
\begin{equation}
\dd s^2 = -\dd\tau^2 + \tau^{2p_t}\dd t^2 + \sum_{i=1}^{d-1} \tau^{2p_i}\dd x_i^2 \ .
\end{equation}
Once the transverse exponents are no longer equal, different extremal surfaces can probe different combinations of the anisotropic Kasner data. Volume-filling slices should be sensitive to combinations involving all $p_i$, whereas an HM surface whose entangling normal lies along $x^k$ may instead depend on a combination that excludes $p_k$. More general curvature- or extrinsic-curvature-dependent functionals may resolve still finer features of the anisotropic geometry. This extension is particularly relevant for spatially anisotropic phases, helical backgrounds, and translation-breaking models. Rotating interiors provide another important setting, although they generally require going beyond the diagonal homogeneous Kasner ansatz to include off-diagonal components and angular dependence. Complexity=Anything offers a natural framework for exploring these effects, while analytic hairy black-hole solutions provide controlled laboratories in which generalized complexity, the thermal $a$-function, and translation-breaking matter can be studied together~\cite{Caceres:2022hei, Liu:2022rsy, Arean:2024pzo, Xu:2025edz}.

Beyond generalized complexity and extremal surfaces, time-like entanglement entropy (TEE) offers a qualitatively different probe of the black hole interior. As a single-boundary observable, holographic timelike entanglement has been shown to be sensitive to curvature singularities, competing complex saddles, and distinct causal structures behind the horizon~\cite{Doi:2022iyj,Doi:2023zaf,Li:2022tsv,Anegawa:2024kdj,Guo:2025pru,Afrasiar:2025eam}. It would therefore be useful to derive its large-temporal-width behavior directly in a general Kasner interior and determine whether the real and imaginary parts of the entropy select different combinations of $(p_t,p_s)$. One may also ask whether the existence of the relevant complex or Lorentzian extremal branches imposes additional Kasner bounds, and whether these bounds are related to bulk energy conditions in a manner similar to the probes studied here. Time-like entanglement first law constructions, including their higher-curvature extensions, may provide a further way to constrain perturbations of Kasner interiors~\cite{Li:2025tud, Xiao:2026nvu}.

A further extension is to relax the assumption that the interior settles into a single terminal Kasner regime. Matter interactions, Kasner transitions, and higher-derivative corrections can generate several successive Kasner epochs or higher-curvature eons before the terminal singularity, and different holographic probes need not be sensitive to the same epoch~\cite{Caceres:2024edr}. Different boundary observables can retain information about different aspects of the near-singularity geometry; for example, recent work shows that Kasner exponents and Kasner transitions can be extracted from the large-overtone quasinormal spectrum of asymptotically AdS black holes~\cite{Hartnoll:2026vhu}. Related recent work showed that distinct high-overtone quasinormal-mode response channels can independently reconstruct the temporal and spatial Kasner exponents without imposing the Kasner constraint~\cite{Xiao:2026pir}. Terminal quantities such as the CA action and the thermal $a$-function are naturally most sensitive to the deepest accessible regime, whereas the critical surfaces controlling CV and HM may remain localized in an earlier epoch. Heavy-operator correlators provide yet another form of selectivity: in known multi-eon examples, a positive $p_t$ can prevent neutral space-like geodesics from penetrating into deeper regimes, while epochs with $p_t\leqslant0$ can remain accessible to the corresponding geodesic branch~\cite{Frenkel:2020ysx}. An eon transition may therefore create, merge, or remove extrema of the CV and HM control functions while simultaneously changing which parts of the interior are visible to boundary correlators. A natural next step is to derive matching rules for these different probe functions across successive epochs and to determine how the energy-condition analysis is modified once higher-curvature terms control the local gravitational dynamics.
  
\begin{acknowledgments}
This work is supported by the Natural Science Foundation of China under Grant No. 12375051 and Tianjin University Self-Innovation Fund Extreme Basic Research Project Grant No. 2025XJ22-0014 and 2025XJ21-0007.
\end{acknowledgments}

\appendix

\section{Details of the Complexity=Action Bulk Criterion}
\label{app:CABulk}

\subsection{WdW causal structure}

To understand the near-singularity behavior of complexity, it is crucial to determine whether the WdW patch actually intersects the space-like singularity, which depends on the causal structure of the black hole interior\footnote{The classification of black hole interiors based on causal structure has been explored in various contexts. Pioneering work utilized in-going null geodesics to describe the Penrose diagram's shape~\cite{Fidkowski:2003nf}. This has recently been extended to CA complexity, distinguishing between Type-I and Type-II spacetimes~\cite{An:2022lvo, Caceres:2022smh, Auzzi:2022bfd}.}. To characterize this, we consider an in-falling null sheet departing from the asymptotic boundary at coordinate time $t=0$. By transforming to the Eddington-Finkelstein coordinate $u = t - z^*(z)$ with $\dd z^* = \dd z \, f^{-1}(z)\e^{\chi(z)/2}$, we can define a characteristic regularized time $t_0$ at which this null sheet hits the singularity~\cite{An:2022lvo, Caceres:2022smh, Auzzi:2022bfd}:
\begin{equation}
    t_0 = \int^{\infty}_0 \dd z\left(f^{-1}(z)\e^{\chi(z)/2} - \frac{2\e^{\chi(z_h)/2}z_h}{f'(z_h)(z^2-z^2_h)}\right)\ .
\end{equation}
The magnitude and sign of $t_0$ explicitly dictate the geometry of the WdW patch, as dictated in Fig.~\ref{fig:Type-I-II-CA}. The corresponding relations between the null boundaries, the symmetric anchoring time $t_B$, and the critical times at which the WdW patch changes its causal arrangement follow directly from the equations for the ingoing and outgoing null sheets. 
Assuming symmetric boundary anchoring times $t_R=-t_L\equiv t_B$, the sign of $t_0$ determines how the WdW patch encounters the space-like singularity. 

For $t_0>0$, corresponding to a Type-I spacetime, the WdW patch already intersects the singularity at $t_B=0$, as shown in Fig.~\ref{fig:Type-I-II-CA}(a). The limiting Type-I case $t_0=0$ lies at the critical point, and the WdW patch begins to intersect the singularity for any $t_B>0$. For $t_0<0$, corresponding to a Type-II spacetime, the WdW patch initially avoids the singularity. As the boundary time increases, its null boundaries eventually reach the terminal region at the critical time $t_{B,c}=-t_0=|t_0|$. For $t_B>-t_0$, the WdW patch intersects the singularity along an extended spacelike segment, as illustrated in Fig.~\ref{fig:Type-I-II-CA}(b). Thus, irrespective of the causal type, the late-time CA calculation eventually probes the near-singularity region. 

Therefore, regardless of the specific causal type, the late-time evaluation of CA complexity fundamentally requires integrating the bulk action near this singular boundary. To ensure that the complexity remains well-defined and physically meaningful, this near-singularity action contribution must not diverge, as discussed in Sec.~\ref{subsec:CA}.

\subsection{Matter-sector reduction and radial bulk primitive}

We consider the bulk action
\begin{equation}
    \begin{aligned}
        I_\text{bulk} &= \int \dd^{d+1}x \,\sqrt{-g} \, L\ ,\\
        16\pi L &= R + d(d-1) + L_\text{matter}\ .
    \end{aligned}
\end{equation}
Transverse rotational symmetry implies
\begin{equation}
    {T^i}_j=P(z){\delta^i}_j\ .
\end{equation}
To obtain the reduction used in the main text, we impose the additional matter-sector condition Eq.~\eqref{eq:matter_condition}
\begin{equation}
    \partial L_\text{matter}/\partial g^{ij} = 0\ .
\end{equation}
Using the definition of the energy-momentum tensor for minimal coupling matter
\begin{equation}
    \begin{aligned}
        T_{\mu\nu} = -2 \frac{\partial L_\text{matter}}{\partial g^{\mu\nu}} + g_{\mu\nu}L_\text{matter} \ ,
    \end{aligned}
\end{equation}
we get a relation
\begin{equation}\label{eq:matter_relation}
    T_{ij} = g_{ij}L_\text{matter} \ .
\end{equation}
Raising one index we have, ${T^i}_j = g^{ik}g_{kj}L_\text{matter} = {\delta^i}_j L_\text{matter}$. Without applying Einstein summation on $i$, the diagonal component is simply ${T^i}_i = L_\text{matter}$.

The condition $\partial L_{\text{matter}}/\partial g^{ij}=0$, and hence the relation in Eq.~\eqref{eq:matter_relation}, applies to many standard homogeneous holographic backgrounds. For an Einstein-scalar theory with 
\[
    L_{\text{matter}}=-\frac{1}{2}(\partial\phi)^2-V(\phi)\ ,
\]
a purely radial profile $\phi=\phi(z)$ gives $(\partial\phi)^2=g^{zz}(\phi')^2$, so the matter Lagrangian contains no transverse inverse metric. The same is true in Einstein-Maxwell-dilaton models 
\[
    L_\text{matter} = -\frac{1}{2}(\partial\phi)^2 - V(\phi) - \frac{1}{4} Z(\phi)F^2\ ,
\]
with $\phi=\phi(z)$ and $A=A_t(z)\dd t$, for which both the scalar and electric sectors involve only the $t$- and $z$-components of the inverse metric.
Standard $s$-wave holographic superconductors provide another example. For 
\[
    L_\text{matter} = -\frac{1}{4}F^2 - |D\Phi|^2 - m^2|\Phi|^2\ ,
\]
with $\Phi=\phi(z)$ and $A=A_t(z)\dd t$, the transverse covariant derivatives and field strengths vanish, $D_i\Phi=0$ and $F_{i\mu}=0$. The condition also extends to nonlinear electrodynamics and DBI-type models. For example, if 
\[
    L_\text{matter} = \mathcal{K}(\phi,X) - V(\phi)\ ,
\]
with gauge invariant $X = -\frac{1}{4}F^2$, the purely electric ansatz ensures $X = -\frac{1}{2}g^{zz}g^{tt}(A_t')^2$. Since $X$ is independent of $g^{ij}$, the condition holds regardless of the non-linearity of $\mathcal{K}$.

By contrast, the condition need not hold when the matter sector contains active transverse components. Magnetic or dyonic branes with  $F_{ij}\neq 0$, linear-axion momentum-relaxation models, and $p$-wave superconductors provide familiar examples. Such configurations may still preserve macroscopic isotropy, but their matter Lagrangians depend explicitly on $g^{ij}$, and the transverse pressure $P(z)$ in main text is then no longer equal to $L_{\text{matter}}$.

Substituting this transverse component~\eqref{eq:matter_relation} into the mixed Einstein equations ${G^{\mu}}_{\nu} - \frac{d(d-1)}{2}{\delta^{\mu}}_{\nu} = \frac{1}{2}{T^{\mu}}_{\nu}$, we obtain:
\begin{equation}
    \begin{aligned}
        {G^i}_i &= \frac{d(d-1)}{2} + \frac{1}{2}L_\text{matter} \\
        &= \frac{1}{2}\left( 16\pi L - R \right)\ .
    \end{aligned}
\end{equation}
Taking the full trace of the Einstein tensor in $(d+1)$ dimensions yields ${G^{\mu}}_{\mu} = -\frac{d-1}{2}R$. Expanding this trace into temporal, radial, and spatial parts gives:
\begin{equation}
    -\frac{d-1}{2}R = {G^t}_t + {G^z}_z + \frac{d-1}{2}(16\pi L - R)\ .
\end{equation}
The scalar curvature $R$ cancels out completely, allowing us to isolate the total Lagrangian density:
\begin{equation}
    16\pi L = -\frac{2}{d-1}({G^t}_t + {G^z}_z)\ .
\end{equation}
Substituting this result back into the equation allows us to express the total Lagrangian density solely in terms of the $t$- and $z$-components.
\begin{equation}
   16\pi L\sqrt{-g} = -\frac{2\sqrt{-g}}{d-1}({G^t}_t + {G^z}_z)\ .
\end{equation}
By explicitly computing the Einstein tensor components for the metric ansatz~\eqref{eq:AdSmetric}, one can establish an exact total derivative identity:
\begin{equation}
    16\pi L\sqrt{-g} = 2\frac{\dd}{\dd z}\left( \e^{-\chi/2}f(z)z^{-d} \right)\ .
\end{equation}
This equation is the result used in Sec.~\ref{subsec:CA}.

\section{Kasner-Frame Derivation of the Energy-Condition Bounds}\label{app:KasnerEnergy}

\subsection{Einstein tensor of a diagonal Bianchi-I metric}

In Sec.~\ref{sec:EnergyCondition}, we derived the near-singularity energy-condition bounds directly from the metric functions $f(z)$ and $\chi(z)$. In this appendix, we give an independent derivation in the orthonormal frame of a general Kasner geometry. This form makes the physical meaning of the relevant null and dominant energy conditions transparent and provides a short geometric proof that the NEC and DEC imply the criteria summarized in Talbe~\ref{tab:energy-sufficient-criteria}. Standard treatments of Bianchi cosmologies can be found in Refs.~\cite{Ellis:1968vb, Ryan:1975jw}.

Consider the $(d+1)$-dimensional diagonal Kasner metric parameterized by proper time $\tau$:
\begin{equation}\label{eq:app-Kasner}
    \dd s^2 = -\dd\tau^2 + \sum_{A=1}^{d} a_A^2(\tau)(\dd x^A)^2\ .
\end{equation} 
To align this general formalism with the holographic black hole interior~\eqref{eq:Kasner} studied in the main text, we identify the time-like proper-time coordinate $\tau$ with the proper-time variable introduced in Sec.~\ref{sec:Kasner}, which is related asymptotically to the original radial coordinate $z$ by $\tau\propto z^{-(a+1)}$. We divide the spatial directions $x^A$ into the boundary time direction $t$ and the transverse directions $x^i$ ($i = 1, \dots, d-1$).
We define the directional Hubble parameters
\begin{equation}
    H_A \equiv \frac{\dot a_A}{a_A}\ , \qquad \Theta \equiv \sum_{A=1}^{d}H_A\ ,
\end{equation}
where a dot denotes differentiation with respect to $\tau$. The time-time and $AA$ components of the Ricci tensor are
\begin{equation}\label{eq:app-RTensor-general}
    R_{\tau\tau} = -\dot\Theta - \sum_AH_A^2\ , \qquad R_{AA} = a_A^2 \left( \dot H_A+H_A\Theta \right)\ .
\end{equation}
Contracting these expressions gives Ricci scalar
\begin{equation}\label{eq:app-R-general}
    R  = 2\dot\Theta+\Theta^2+\sum_A H_A^2\ .
\end{equation}

We now specialize to a power-law regime $a_A(\tau) = \tau^{p_A}$. Then $H_A = \frac{p_A}{\tau}$, $\dot H_A = -\frac{p_A}{\tau^2}$.
It is convenient to introduce the sum of exponents and the sum of their squares:
\begin{equation}\label{eq:app-SQ}
    S \equiv \sum_A p_A\ ,\qquad Q\equiv\sum_A p_A^2\ .
\end{equation}
Evaluating Eqs.~\eqref{eq:app-RTensor-general} and~\eqref{eq:app-R-general} in the orthonormal frame $e_{\hat{0}} = \partial_\tau$ and $e_{\hat{A}} = a_A^{-1}\partial_{x^A}$, we find:
\begin{equation}
    R_{\hat{0}\hat{0}} = \frac{S-Q}{\tau^2}\ , \qquad R_{\hat{A}\hat{A}} = \frac{p_A(S-1)}{\tau^2}\ ,\qquad R = \frac{Q+S^2-2S}{\tau^2}\ .
\end{equation}
The components of the Einstein tensor $G_{\mu\nu} = R_{\mu\nu} - \frac{1}{2}g_{\mu\nu}R$ with $\mu,\nu =\hat{0},\hat{A}$ are therefore
\begin{equation}\label{eq:app-G00-AA}
    \begin{aligned}
        G_{\hat{0}\hat{0}} &= \frac{S^2-Q}{2\tau^2}\ ,\\
        G_{\hat{A}\hat{A}} &= \frac{1}{\tau^2} \left[ p_A(S-1) - \frac{1}{2} \left( Q+S^2-2S \right) \right]\ .
    \end{aligned}
\end{equation}
As a check, the nontrivial vacuum Kasner relations $S=1$ and $Q=1$ make every component in Eq.~\eqref{eq:app-G00-AA} vanish.

Explicitly, the Kretschmann scalar $K=R_{\mu\nu\rho\sigma} R^{\mu\nu\rho\sigma}$ is given by
\[
    K = \frac{4}{\tau^4}\left[\sum_A p_A^2(p_A-1)^2+\sum_{A<B}p_A^2 p_B^2\right]\ .
\]
Evaluated on our ansatz metric this yields
\begin{equation}
    K=\frac{4}{\tau^4}\left[p_t^2(p_t-1)^2+(d-1)p_s^2(p_s-1)^2+(d-1)p_t^2p_s^2+\frac{(d-1)(d-2)}{2}p_s^4\right]\ .
\end{equation}
Since $d \geqslant 3$ and $p_s>0$, the bracketed expression is positive. This confirms that the curvature invariant diverges as $\tau \to 0$, consistent with the space-like singularity discussed in the main text.

\subsection{Energy density and principal pressures}

Using the Einstein equation $G_{\mu\nu}+\Lambda g_{\mu\nu}=\frac{1}{2}T_{\mu\nu}$, it is useful to express these contractions in terms of the local energy density and principal pressures. Since the radial direction is time-like inside the horizon, we define
\begin{equation}
    \rho\equiv T_{\hat 0\hat 0}\ , \qquad P_t\equiv T_{\hat 1\hat 1}\ ,
    \qquad P_s\equiv T_{\hat i\hat i}\ ,
\end{equation}
where no sum over $i$ is implied. Constructing the null vectors $l_{(\tau, t)} = \frac{1}{\sqrt{2}}(e_{\hat{0}} + e_{\hat{t}})$ and $k_{(\tau, t)} = \frac{1}{\sqrt{2}}(e_{\hat{0}} - e_{\hat{t}})$ in the $\tau-t$ plane, alongside their transverse counterparts $l_{(\tau, x^i)} = \frac{1}{\sqrt{2}}(e_{\hat{0}} + e_{\hat{x}^i})$ and $k_{(\tau, x^i)} = \frac{1}{\sqrt{2}}(e_{\hat{0}} - e_{\hat{x}^i})$ in the $\tau-x^i$ planes, the relevant classical energy conditions elegantly reduce to linear combinations of the local density and pressures:
\begin{equation}
    \begin{aligned}
        (\tau,t)\text{-NEC}\quad \,T_{\mu\nu}l^\mu_{(\tau,t)}l^\nu_{(\tau,t)} \geqslant 0\,: & \quad \rho+P_t \geqslant 0\ , \\
        (\tau,x^i)\text{-NEC}\quad \,T_{\mu\nu}l^\mu_{(\tau,x^i)}l^\nu_{(\tau,x^i)} \geqslant 0\,: & \quad \rho+P_s \geqslant 0\ , \\
        (\tau,t)\text{-DEC}\quad \,T_{\mu\nu}l^\mu_{(\tau,t)}k^\nu_{(\tau,t)} \geqslant 0\,: & \quad \rho-P_t \geqslant 0\ .
    \end{aligned}
\end{equation}

Because the cosmological constant $\Lambda$ cancels out in the null combinations $\rho + P_A$, Eq.~\eqref{eq:app-G00-AA} directly gives
\begin{equation}\label{eq:app-general-NEC}
    \frac{1}{2}(\rho+P_A) = G_{\hat0\hat0} + G_{\hat A\hat A} = \frac{ S-Q+p_A(S-1) }{\tau^2}\ .
\end{equation}
We now specialize to the transversely isotropic Kasner metric used in the main text, where $p_t$ is the longitudinal exponent and $p_s$ applies to the $(d-1)$ transverse directions. Consequently, the summation terms become:
\begin{equation}
    S=p_t+(d-1)p_s\ ,\qquad Q=p_t^2+(d-1)p_s^2\ .
\end{equation}

For the longitudinal direction inherited from the boundary time ($A = t$), Eq.~\eqref{eq:app-general-NEC} yields
\begin{equation}\label{eq:app-longitudinal-NEC}
    \frac{1}{2}(\rho+P_t) = \frac{ S-Q+p_t(S-1) }{\tau^2} = \frac{ (d-1)p_s(1+p_t-p_s) }{\tau^2}\ .
\end{equation}
Given the physical constraint $p_s > 0$, the longitudinal $(\tau,t)$-NEC necessitates $\rho + P_t \geqslant 0$, which simplifies immediately to:
\begin{equation}
\label{eq:app-longitudinal-bound}
    p_t\geqslant p_s-1\ .
\end{equation}
Hence Eq.~\eqref{eq:app-longitudinal-bound} is in agreement with the $f,\chi$ derivation~\eqref{eq:NECzt} in the main text.

For any transverse direction ($A = x^i$), substituting $p_A = p_s$ into Eq.~\eqref{eq:app-general-NEC} instead gives
\begin{equation}\label{eq:app-transverse-NEC}
    \frac{1}{2}(\rho+P_s) = \frac{ S-Q+p_s(S-1) }{\tau^2} = \frac{ -p_t^2 +(p_s+1)p_t +(d-2)p_s }{\tau^2}\ .
\end{equation}
Enforcing the transverse $(\tau,x^i)$-NEC ($\rho + P_s \geqslant 0$) therefore yields the quadratic constraint:
\begin{equation}
\label{eq:app-transverse-bound}
    p_t^2-(p_s+1)p_t-(d-2)p_s \leqslant 0\ ,
\end{equation}
consistent with the main-text derivation~\eqref{eq:NECzx}.

Finally, consider the dominant energy condition $\rho-P_t\geqslant0$ in the longitudinal sector. From the Einstein equations, this combination does not cancel the cosmological constant:
\begin{equation}\label{eq:app-longitudinal-DEC}
    \begin{aligned}
        \frac{1}{2}(\rho-P_t) &= G_{\hat0\hat0} - G_{\hat1\hat1} - 2\Lambda \\
        &= \frac{ (d-1)p_s \left[ p_t+(d-1)p_s-1 \right] }{\tau^2} - 2\Lambda\ .
    \end{aligned}
\end{equation}
For unit AdS radius, $-2\Lambda=d(d-1)$, and because the $\tau^{-2}$ term dominates near the singularity, the DEC cannot hold if $p_t+(d-1)p_s<1$. Deep in the Kasner interior ($\tau \to 0$), the $\tau^{-2}$ term dominates the finite constant $\Lambda$. If the coefficient of the divergent term is negative ($p_t+(d-1)p_s<1$), the expression will plunge to $-\infty$, violating the DEC. Therefore, maintaining $\rho - P_t \geqslant 0$ requires the leading coefficient to be non-negative:
\begin{equation}
\label{eq:app-DEC-bound}
    p_t+(d-1)p_s\geqslant 1\ .
\end{equation}
This agrees with the condition~\eqref{eq:DECzt} obtained from $f$ and $\chi$.

\bibliographystyle{JHEP}

\bibliography{Kasner-NEC}

\end{document}